\documentclass[print, amsmath,amssymb, aps,prc,nofootinbib,notitlepage,superscriptaddress]{revtex4-2}

\usepackage[usenames,dvipsnames]{xcolor}
\usepackage{graphicx}
\usepackage{bm}
\usepackage[utf8]{inputenc}
\usepackage{amsmath}
\usepackage{soul} 
\setstcolor{blue} 
\usepackage{bbold} 

\newcommand{\nn}{\nonumber}
\newcommand{\q}{\quad}

\renewcommand{\vec}[1]{{\bf #1}}

\newcommand{\etal}{\emph{et al.}}

\newcommand{\bg}{\boldsymbol{\mathbf g}}

\newcommand{\bp}{\boldsymbol{\mathbf p}}
\newcommand{\br}{\boldsymbol{\mathbf r}}
\newcommand{\brp}{\boldsymbol{\mathbf r'}}
\newcommand{\bs}{\boldsymbol{\mathbf s}}

\newcommand{\bC}{\boldsymbol{\mathbf C}}
\newcommand{\bJ}{\boldsymbol{\mathbf J}}
\newcommand{\bR}{\boldsymbol{\mathbf R}}
\newcommand{\bV}{\boldsymbol{\mathbf V}}
\newcommand{\bW}{\boldsymbol{\mathbf W}}

\newcommand{\CE}{\mathcal{E}}
\newcommand{\CH}{\mathcal{H}}

\newcommand{\bc}{\begin{center}}
\newcommand{\ec}{\end{center}}
\newcommand{\be}{\begin{equation}}
\newcommand{\ee}{\end{equation}}
\newcommand{\bqr}{\begin{eqnarray}}
\newcommand{\eqr}{\end{eqnarray}}
\newcommand{\bi}{\begin{itemize}}
\newcommand{\ei}{\end{itemize}}
\newcommand{\bwt}{\begin{widetext}}
\newcommand{\ewt}{\end{widetext}}
\newcommand{\bsub}{\begin{subequations}}
\newcommand{\esub}{\end{subequations}}
\newcommand{\bnum}{\begin{enumerate}}
\newcommand{\enum}{\end{enumerate}}
\newcommand{\tr}{\rm Tr}

\newcommand{\vnabla}{\boldsymbol{\mathbf\nabla}}
\newcommand{\vsigma}{\boldsymbol{\mathbf\sigma}}

\newcommand{\citeqdot}[1]{Eq.~(\ref{#1})}

\newcommand{\citeRefdot}[1]{Ref.\cite{#1}}
\newcommand{\citeRefsdot}[1]{Refs.\cite{#1}}

\newcommand{\citeFigure}[1]{Figure~\ref{#1}}

\newcommand{\citeAppendix}[1]{Appendix~\ref{#1}}

\newcommand{\citeTable}[1]{Table~\ref{#1}}

\begin{document}

\title{Surface energy coefficient of a N2LO Skyrme energy functional :
       a semiclassical Extended Thomas-Fermi approach}

\author{P. Proust}
\email{p.proust@ip2i.in2p3.fr}
\affiliation{Universit\'e de Lyon, F-69003 Lyon, France \\
Institut de Physique des 2 Infinis, CNRS-IN2P3,  \\ 
UMR 5822, Universit\'e Lyon 1, F-69622 Villeurbanne, France}
						
\author{Y. Lallouet}
\affiliation{Lyc\'ee Malherbe, 14, Avenue Albert Sorel, 14000 Caen}

\author{D. Davesne}
\email{d.davesne@ip2i.in2p3.fr}
\affiliation{Universit\'e de Lyon, F-69003 Lyon, France \\
Institut de Physique des 2 Infinis, CNRS-IN2P3,  \\ 
UMR 5822, Universit\'e Lyon 1, F-69622 Villeurbanne, France}

\author{J. Meyer}
\email{jmeyer@ip2i.in2p3.fr}
\affiliation{Universit\'e de Lyon, F-69003 Lyon, France \\
Institut de Physique des 2 Infinis, CNRS-IN2P3,  \\ 
UMR 5822, Universit\'e Lyon 1, F-69622 Villeurbanne, France}

\begin{abstract}
We generalize to N2LO Skyrme functionals the semi-classical approach of Grammaticos and Voros~\cite{Gra78,Gra79}
in order to calculate the Extended Thomas Fermi expressions of the new densities and currents appearing at the N2LO level.
Within a one dimensional symmetric semi infinite nuclear matter model and using a simple Fermi-like density profile,
we obtain an easy-to-use formula for the surface energy including the contributions of the central, density-dependent 
and spin-orbit terms up to $\hbar^2$.
Such a formula can be easily used as a first attempt to constrain the surface properties of new N2LO Skyrme parametrizations. 
The N2LO parametrization tested in this paper is shown to exhibit a shift (compared to a full Hartree-Fock calculation) which is
quantitatively similar to the one obtained with the traditional Skyrme parametrizations. 
\end{abstract}

\date{\today}

\maketitle

%
%
\section{Introduction}
%
%
Microscopic self-consistent models based on the mean field approximation have demonstrated 
their usefulness to describe precisely a wide range of nuclear properties such as binding energies, 
radii, separation energies, energy spectra or also fission barriers of heavy nuclei.
At this time, they even allow rather pertinent analysis of spectroscopy of superheavy nuclei for which 
there exists a lot of recent experimental results.
All these models use as a basic ingredient an effective interaction containing parameters
which must be determined from a panel of pseudo and experimental data within a precise protocol.   
Among the pseudo data used in the fitting procedure, properties of infinite matter are usually incorporated 
but those of semi-infinite nuclear matter are often neglected because of the required computational time. However, 
the surface energy has been recognized in the early years as one of the fundamental 
properties of nuclei and thus entered the liquid drop formula even in its simplest form. 
In this particular formula, its role is to keep a nucleus spherical (against Coulomb repulsion for instance) 
but more generally, its physical meaning is related to the shape of a nucleus and its ability to deform. 
From a fundamental point of view, since this last property is controlled by the compressibility for small 
density oscillations, both quantities are related~\cite{Myers90,Sto80}. 
From a practical point of view, the surface energy naturally contributes to the deformation properties 
of nuclei~\cite{Bar82,Ryssens19} and thus in fusion reactions~\cite{Ste19,Sal11,Dut10,Gha15,Gol13}, 
to the determination of fission barriers~\cite{Tos13,Gou05} or in cluster decay~\cite{Raj17} 
and neutron-star matter~\cite{Rav72}.
It is therefore of the utmost importance to determine precisely the surface energy and each of the 
aforementionned constraints could be used to fine tune the parameters of an effective 
interaction~\cite{Bar82,Gor07,Nik10}. 
Historically, this was not the case and the huge difference between the result obtained 
with a Skyrme parametrization and the experimental 
fission barrier height of $^{240}$Pu induced the adjustment of SkM*~\cite{Bar82}. This parametrization  
is a modified version of SkM~\cite{Treiner80} fine tuned in such a way that the height 
of a semi classical calculation of the fission barrier~\cite{Bra85} is well reproduced  without 
any modification of infinite nuclear matter isoscalar properties. 
The parametrization SkM* is still quite commonly used for the description of fission phenomena~\cite{Schunck14}.

Since neither the fission barriers neither a full Hartree-Fock (HF) calculation of the surface energy~\cite{Cot78} 
can be directly used in a fitting protocol because of the numerical cost, semi-classical methods and especially 
the Extended Thomas Fermi (ETF) expansion method~\cite{Brack_Bhaduri_book,Bra76,Byron_thesis,Gra78,Gra79,Bra85} 
were developed as an alternative. This kind of approach is of great interest when one is concerned 
by an approximation of the total energy of a fermionic system, let us say a nucleus made with neutrons and protons, 
but not by an approximation of the total wave function. 
Generally speaking, semi classical methods are mainly based on the Strutinsky theorem, such that one can split 
the Hartree-Fock energy of a system of fermions in two parts: 
an important smoothly varying part which can be roughly described by the liquid drop model, 
plus a smaller, but not smooth, part purely related to shell effects~\cite{Strut67,Strut68,Brack73,Brack75,Brack81}. 
Semi-classical expansions are actually known to give an accurate description of the smooth part of the energy. 
Practically speaking, they consist of an expansion in powers of $\hbar$ of the density matrix which allows us to 
express each of the densities and currents appearing in the energy density functional as functions of the 
Fermi modulus $k_F$, the effective mass field $f$ and their spatial derivatives.
Then, by eliminating $k_F$, one can express the energy with respect to the density and it's spatial derivatives alone. 
The great advantage comes from the variational principle with respect to the individual wave functions 
which is replaced by a simpler variational principle with respect to the density profile itself. 
This leads to an easy-to-use formula which reduces drastically the computer time when compared to microscopic HF calculations. 
Two methods have been used in the past to obtain an ETF expansion. First an expansion in powers of $\hbar$ 
of the Bloch density~\cite{Brack_Bhaduri_book,Bra76,Byron_thesis,Bra85} and second the Wigner-Kirkwood transform 
of a one-body hamiltonian~\cite{Gra78,Gra79}. 
A direct and recent application of this approach can be found in Jodon \etal~\cite{Jod16} and in Ryssens \etal~\cite{Ryssens19} 
where the authors examined the correlation between the surface energy and the deformation properties of heavy nuclei. 
In Jodon \etal~\cite{Jod16} in particular, the authors revisited a very simple ETF expansion obtained 
by Krivine and Treiner~\cite{Krivine79,Treiner86} in order to obtain a simple formula for the surface energy 
and thus to obtain an easy-to-implement constraint in the fitting procedure which leads to the so-called SLy5s 
parametrizations~\cite{Jod16}. 
In this work, we extend the method developed by Grammaticos and Voros~\cite{Gra78,Gra79} to Skyrme energy density 
functional (EDF) containing 4th gradient terms (named N2LO) which are expected to improve the calculated properties 
of nuclei~\cite{Becker17}.
Temperature effects, relativistic models as well as the possibility of superfluidity are not been considered here. 
%
%
%
The paper is organized as follows. Section II recalls the N2LO Skyrme EDF considered here.
Section III generalizes the results obtained in \cite{Gra78,Gra79} for the new densities appearing at the N2LO order.
Section IV uses the semi-classical ETF results of the third section to calculate the surface energy
of a N2LO Skyrme functional within a 1D semi-infinite nuclear matter model.
Section V presents the discussion of our results.
%
%
\section{The Skyrme Energy Density Functional at the N2LO order}
\label{sect:EDF_Skyrme}
%
%
%
%
\subsection{Local densities and currents}
\label{sect:densities}
%
%
The interaction used in this paper is the so-called N2LO 
extension~\cite{Carlsson08,Carlsson10,Raimondi11a,Raimondi11b} of the traditional Skyrme interaction. 
It implies gradients up to fourth order. 
In \citeRefdot{Ryssens21}, the authors considered specific densities more suited to 3D HF numerical 
calculations but for this formal work, we keep here the more compact notations for densities introduced in~\cite{Davesne13}. 
The density matrix elements (the $q$-index is for the n/p representation) in direct and spin spaces reads: 
\bqr
\langle \br \sigma | \hat{\rho}_q | \br' \sigma' \rangle 
        & = & \rho_q( \br \sigma, \br' \sigma' ) 
       \, = \, \frac{1}{2} \, \rho_q ( \br, \br' ) \, \delta_{\sigma\sigma'}
       \, + \, \frac{1}{2} \, \bs_q ( \br, \br' ) \, \langle \sigma | \hat{\vsigma} | \sigma' \rangle  
			 \, = \, \langle \br \sigma | \frac{\hat{\rho}_{q,\br} \otimes \hat{\mathbb{1}}_{\sigma} 
			                      \, + \, \hat{\bs}_{q,\br} \otimes \hat{\vsigma}}{2} | \br' \sigma' \rangle      \q ,
\eqr
where
\bsub
\begin{eqnarray}
\hat{\rho}_{q,\br} & = & \tr_\sigma (\hat{\rho}_q)                  \q , \\
\hat{\bs}_{q,\br}  & = & \tr_\sigma (\hat{\rho}_q \hat{\vsigma})    \q .
\end{eqnarray}
\esub
We introduce now the useful local densities as:
\bsub
\begin{eqnarray}
\label{eq:locdensities:rho}
\rho_q (\br) & = & \, \rho_q (\br,\br^\prime) \big|_{\br = \br^\prime}  \q , \\
\label{eq:locdensities:s}
\bs_q (\br)  & = & \, \bs_q  (\br,\br^\prime) \big|_{\br = \br^\prime}  \q , \\
             &   &                                                       \nn \\
\label{eq:locdensities:tau}
\tau_q (\br) & = & \, \vnabla \cdot \vnabla' \; \rho_q (\br,\brp) \big|_{\br=\brp}      \q , \\
\label{eq:locdensities:J}
\tau_{q, \mu\nu}^{(s)} (\br) & = & \, \frac{1}{2} \, \big( \nabla_\mu\nabla_\nu' + \nabla_\nu\nabla_\mu' \big) 
                                                  \, \rho_q (\br,\brp) \big|_{\br=\brp} \q , \\
Q_q (\br) & = & \, \Delta \Delta' \rho_q(\br,\brp) \big|_{\br=\brp}                     \q , \\
          &   &                                                                          \nn \\
J_{q , \mu \nu} (\br) & = & \frac{1}{2i} \, \big( \nabla_\mu - \nabla_\mu^\prime \big) 
                                         \, s_{\nu,q} (\br,\br') \big|_{\br = \br'}     \q , \\
C_{q, \mu \nu} (\br)  & = & \, \frac{1}{2i} 
		                        \, \big( \, \Delta^{\prime} \nabla_\mu - \Delta \nabla_\mu^{\prime} \, \big)
														\, s_\nu ( \br, \br^{\prime})  \big|_{\br=\br^{\prime}}     \q .
\eqr
\esub
Time reversal invariance is assumed which restrict our work to even-even nuclei or unpolarized infinite systems. 
Using this assumption, the time odd part of the Skyrme EDF vanishes so that the corresponding densities do not appear in the list above.
The treatment of the time odd part of the NLO part of the Skyrme EDF has been made 
by Grammaticos and Voros in~\cite{Gra78}. It requires the definition of the corresponding currents but does not 
imply any specific issues and will be treated in future developments.
  
%
%
\subsection{The Skyrme EDF}
\label{sect:EDF}
%
%
We consider in this paper a Skyrme pseudo potential containing a 2-body 
central (C) term and a spin-orbit (SO) term up to NLO order (see~\citeRefdot{Bender03} for notations) plus a central N2LO component as developed in~\cite{Becker17}. 
Let us note that the tensor (T) part of two body pseudo-potentials as developed in~\cite{papI} could be also
introduced without major changes since the spin-gradient coupling present in the C part already generates 
a tensor component in the Skyrme EDF (see the expression of $\CH^{\rm NLO}_{\rm T}$, \citeqdot{eq:full_EDF} below).
Similarly, the central three body part developed in~\cite{Sadoudi_thesis,Sadoudi13} could be also taken into 
account by a modification of the concerned one body fields. 

Finally, we can write the functional as (the $q=n/p$ refers to the neutron/proton representation 
while the $t=0,1$ index refers to the isoscalar/isovector representation): 

\bsub
\bqr
\label{eq:full_EDF}
\CH  & = & \, \CH^{\rm LO}_{\rm C} 
      \, + \, \CH^{\rm NLO}_{\rm C}  \, + \, \CH^{\rm NLO}_{\rm SO} \, + \, \CH^{\rm NLO}_{\rm T} 
			\, + \, \CH^{\rm N2LO}_{\rm C} \, + \, \CH^{\rm N2LO}_{\rm T}                                    \q , \\
		 &   &                                                                                              \nn \\	
\CH^{\rm LO}_{\rm C}  & = & \, \sum_{t=0,1} \, C_t^{\rho} \left[ \rho_0 \right] \, \rho_t^2 
                     \, = \, \sum_{t=0,1} \, \left[ \, C_t^{\rho} 
		                                  \, + \, C_t^{\rho^\gamma} \, \rho_0^\gamma \right] \, \rho_t^2   \q , \\ 
\CH^{\rm NLO}_{\rm C}  & = & \, \sum_{t=0,1} \, C_t^{\Delta \rho} \, \rho_t \, \Delta \rho _t                         
	                      \, + \, C_t^\tau \, \rho_t \, \tau_t                                           \q , \\ 
\CH^{\rm NLO}_{\rm T}  & = & \, - \, \frac{1}{2} \, \sum_{t=0,1} \, C_t^T \, \bJ_{t}^2                 \q , \\
\CH^{\rm NLO}_{\rm SO} & = & \, \sum_{t=0,1} \, C_t^{\nabla J} 
                                             \, \rho_t \, \left( \vnabla \cdot \bJ_t \right)           \q , \\
\CH^{\rm N2LO}_{\rm C} & = & \, \sum_{t=0,1} \, C_t^{(\Delta \rho)^2} \, \left( \Delta \rho_t \right)^2    
                                        \, + \, C_t^{M \rho} 
																				     \, \Big[ \, \rho_t \, Q_t \, + \, \tau_t^2                     
                                                 \, + \, 2 \, \left( \, \tau^{(s)}_{t,\mu \nu} \tau^{(s)}_{t,\mu \nu} 
                                                                \, - \, \tau^{(s)}_{t,\mu \nu} \, \nabla_{\mu} \nabla_{\nu} \rho  
																													 \, \right) \, \Big]                         \q , \\                    
\CH^{\rm N2LO}_{\rm T} & = & \, \sum_{t=0,1} 
                             \, C_t^{MT} 
                             \, \Big[ \, 2 \, \bJ_{t} \cdot \bC_{t}   
                                 \, - \, \frac{1}{2} \bJ_{t} \cdot \Delta \bJ_{t} \, \Big]             \q ,                    
\eqr
\esub
where $\bJ_t$ and $\bC_t$ are the vector components of $J_{t, \mu \nu}$ and $C_{t,\mu \nu}$ respectively 
and where the repeated $\mu,\nu$ indexes are summed over the cartesian coordinates. 
At this stage, we made a second assumption which induced the reduction of the 
tensor currents to their vector part. 
This occurs in spherical nuclei and also in semi infinite nuclear matter as we will see in a further section.

%
%
\subsection{The one body hamiltonian}
\label{sect:fields}
%
%
The one body hamiltonian associated to the above N2LO EDF is given by:
\begin{eqnarray}   
\label{eq:hfield}
					        h_q (\br \sigma, \br' \sigma') 
					& = & \, \langle \br \sigma | \hat{h}_q | \br' \sigma' \rangle 
		     \, = \, \frac{\delta E}{\delta \rho_q (\br' \sigma', \br \sigma)} 
				 \, = \, \frac{\delta }{\delta \rho_q (\br' \sigma', \br \sigma)} 
				      \, \int_{\mathbb{R}^3} \, d^3\br^{\prime \prime} \, \CH (\br^{\prime \prime})           \nn \\ 
				  & = & \, U_q (\br) \, \delta(\br - \br^\prime) \, \delta_{\sigma \sigma'}  
					 \, - \, \nabla_{\mu} \bigg[ \, B_{q,\mu\nu} (\br) \nabla_\nu \delta(\br - \br^\prime) 
														 \, \bigg] \, \delta_{\sigma \sigma'} 
					 \, + \, \Delta \bigg[ \, D_q (\br) \Delta \delta(\br - \br^\prime) 
								       \, \bigg] \, \delta_{\sigma \sigma'}                          \nn \\
                                    &   & 
					 \, + \, i \, \bW_q (\br) \cdot \bigg( \langle \sigma | \hat{\vsigma} | \sigma' \rangle 
												\times \vnabla \delta(\br - \br^\prime) \bigg) 
					 \, + \, i \, \vnabla \cdot \bigg[ \, \left[ 
										\, \bV_q (\br) \cdot \left( \langle \sigma | \hat{\vsigma} | \sigma' \rangle \times \vnabla \right) 
																					   \, \right] \, \vnabla \delta(\br - \br^\prime) \bigg] \q , 
\end{eqnarray}
where the fields can be separated in their NLO+N2LO components as        

\bsub
\bqr
      U_q (\br)               & = & \, U_q^{\text{NLO}} (\br) 
                               \, + \, U_q^{\text{N2LO}} (\br)                 \q , \\  
      B_{q, \mu\nu} (\br) & = & \, B_q^{\text{NLO}} (\br) \, \delta_{\mu\nu}
                           \, + \, B_{q, \mu\nu}^{\text{N2LO}} (\br)           \q , \\  
			          D_q (\br) & = & \, D_q^{\text{N2LO}} (\br)                     \q , \\  
              \bW_q (\br) & = & \, \bW_q^{\text{NLO}} (\br) 
                           \, + \, \bW_q^{\text{N2LO}} (\br)                   \q , \\
              \bV_q (\br) & = & \, \bV_q^{\text{N2LO}} (\br)                   \q .
\eqr
\esub
The expressions of all the above fields in terms of densities and currents are given in~\citeAppendix{app:fields}.
At the N2LO order two new fields appear, i.e. $D_q (\br)$ and $\bV_q (\br)$.
Moreover, $B_q (\br)$ field is no longer a scalar (as for NLO) but now becomes a symmetric rank 2 tensor with a
non-zero contribution of its scalar part and its deviator (i.e. symmetric and traceless) part. 

%
%
\section{The semi classical Extended Thomas Fermi approximation}
\label{sec:ETF_GV}
%


\subsection{The Wigner-Kirkwood transform}
\label{sec:wigner}


We now focus on the semi-classical expansion of the N2LO functional and follow the notations 
of Grammaticos and Voros~\cite{Gra78,Gra79}. 
The Wigner-Kirkwood transform of an operator $\hat A$ and of a product of operators $\hat A\hat B$ 
are respectively given by \cite{Weyl1927}:
\begin{eqnarray}
A_{\scriptscriptstyle W} (\bR, \bp, \hbar) = \int_{\mathbb{R}^3} d^3 \vec{r} \ \langle \bR -\tfrac{1}{2}\br | \hat A(\hbar) | \bR +\tfrac{1}{2}\br \rangle e^{i \bp \br/\hbar}
\end{eqnarray}
and \cite{Groenewold1946} \footnote{Note the presence of some sign errors in the derivation of the following formula in Ref~\cite{Groenewold1946}. However, the final result is correct.}
\begin{eqnarray}
\label{TWproduit}
(\hat A\hat B)_{\scriptscriptstyle W} = A_{\scriptscriptstyle W}  e^{\frac{1}{2}i\hbar \overleftrightarrow \Lambda} B_{\scriptscriptstyle W} 
\end{eqnarray}
with
\begin{eqnarray}
\overleftrightarrow \Lambda \equiv \sum_i \left( \frac{\overleftarrow \partial}{\partial r_i}\frac{\overrightarrow \partial}{\partial p_i} - \frac{\overleftarrow \partial}{\partial p_i}\frac{\overrightarrow \partial}{\partial r_i} \right)
\end{eqnarray}
where the above operator acts on left or right side depending on the arrows.
The object of interest we want to expand in powers of $\hbar$ is the density matrix which, at zero temperature, reads :
\begin{equation}
\hat{\rho}_q = \Theta (E_{F,q} \, \hat{\mathbb{1}}_{\br \otimes \sigma} - \hat{h}_q)
\end{equation}
where $\hat{h}_q$ is the one-body hamiltonian and $E_{F,q}$ the Fermi energy of species $q$.
Since we consider the case with a spin-orbit term, the Wigner transform of $\hat{h}_q$ can thus be expressed 
as ($\vec x \equiv (\vec{R},\bp)$):
\begin{equation}
(\hat{h}_q)_{\scriptscriptstyle W} (\vec x,\hbar) = H_{q,cl} (\vec{R},\bp) \hat{\mathbb{1}}_\sigma
           + \hbar \, \vec{H_{q,so}} (\vec{R},\bp) \cdot \hat{\vsigma} 
\end{equation}
where $H_{q,cl}$ represents the classical limit (obtained formally when $\hbar$ goes to zero) and $\vec{H_{q,so}}$ 
the spin-orbit part, which enters the expansion as a first-order $\hbar$ correction (see~\citeRefdot{Gra79}). 
Given any analytical functions $F$ of the hamiltonian $\hat{h}_q$, the expansion in the vicinity of its classical value reads:
\begin{eqnarray}
F(\hat h_q) = \sum_n^\infty \frac{1}{n!} F^{(n)}(H_{q,cl}(\vec x)) (\hat h_q - H_{q,cl}(\vec x) \hat{\mathbb{1}}_{\br \otimes \sigma})^n
\end{eqnarray}
so that the Wigner-Kirkwood transform is
\begin{eqnarray}
(F(\hat h_q))_{\scriptscriptstyle W} (\vec x, \hbar) = \sum_n^\infty \frac{1}{n!} F^{(n)}(H_{q,cl}(\vec x)) \hat{\cal G}_{q,n}(\vec x, \hbar)
\end{eqnarray}
with
\begin{eqnarray}
\hat{\cal G}_{q,n}(\vec x, \hbar) = \left.((\hat h_q - H_{q,cl}(\vec x) \hat{\mathbb{1}}_{\br \otimes \sigma})^n)_{\scriptscriptstyle W} (\vec x ', \hbar)\right|_{\vec x ' = \vec x}
\end{eqnarray}
which are universal operators for a given Hamiltonian, i.e. independent of the function $F$. 
At order $\hbar^3$, we recover the results of Grammaticos and Voros~\cite{Gra79} :
\begin{eqnarray}
\hat{\cal G}_{q,0} (\vec{x}, \hbar) & = & \, \hat{\mathbb{1}}_\sigma                               \nonumber \\
\hat{\cal G}_{q,1} (\vec{x}, \hbar) & = & \, \hbar \, \vec{H_{q,so}} \cdot \hat{\vsigma}  \nonumber \\
\hat{\cal G}_{q,2} (\vec{x}, \hbar) & = & \, \frac{\hbar^2}{4} 
                                          \, \big[ \, 4 \, \vec{H_{q,so}}^2 
																					    \, + \, (\partial^2_{x_\mu p_\nu} H_{q,cl}) \, (\partial^2_{x_\nu p_\mu} H_{q,cl}) 
																							\, - \, (\partial^2_{x_\mu x_\nu} H_{q,cl}) \, (\partial^2_{p_\mu p_\nu} H_{q,cl}) 
																					\, \big] \, \hat{\mathbb{1}}_\sigma 
																					\, + \, {\cal O} (\hbar^3)                      \nonumber \\
\hat{\cal G}_{q,3} (\vec{x}, \hbar) & = & \, \frac{\hbar^2}{4} 
                                          \, \big[ \, 2 \, (\partial^2_{x_\mu p_\nu} H_{q,cl}) \, (\partial_{p_\mu} H_{q,cl})
																					              \, (\partial_{x_\nu} H_{q,cl}) 
																							\, - \, (\partial^2_{x_\mu x_\nu} H_{q,cl}) \, (\partial_{p_\mu} H_{q,cl})
																							     \, (\partial_{p_\nu} H_{q,cl})         \nonumber \\
                                    &   & \hskip 6 true cm 
																		          \, - \, (\partial^2_{p_\mu p_\nu} H_{q,cl}) \, (\partial_{x_\mu} H_{q,cl})
																							     \, (\partial_{x_\nu} H_{q,cl}) \, \big] \, \hat{\mathbb{1}}_\sigma 
																					\, + \, {\cal O} (\hbar^3)                      \nn
\end{eqnarray}
We thus immediately obtain for the density matrix :
\begin{eqnarray}
(\hat{\rho_q})_{\scriptscriptstyle W} (\br,\bp) & = & \, \Theta(E_{F,q}-H_{q,cl}(\br,\bp)) \, \hat{\mathbb{1}}_\sigma 
                            \, - \, \sum_{n=1}^\infty \, \frac{\hat{\cal G}_{q,n}(\vec{x}, \hbar)}{n!} \, \delta^{(n-1)} \, ( H_{q,cl}(\br,\bp) - E_{F,q} ) \nn
\end{eqnarray}
and the non zero densities and currents read 
\bqr
     \rho_q (\br) & = & \, \int_{\mathbb{R}^3} \frac{d^3 \vec{p}}{(2\pi\hbar)^3} 
		                    \, (\hat{\rho}_{q,\br})_{\scriptscriptstyle W} (\br,\bp)          \q , \\ 
     \tau_q (\br) & = & \, \frac{1}{4} \, \Delta \, \int_{\mathbb{R}^3} \frac{d^3 \vec{p}}{(2\pi\hbar)^3} 
		                    \, (\hat{\rho}_{q,\br})_{\scriptscriptstyle W} (\br,\bp) 
		               \, + \, \int_{\mathbb{R}^3} \frac{d^3 \vec{p}}{(2\pi\hbar)^3} \, \frac{p^2}{\hbar^2} 
									      \, (\hat{\rho}_{q,\br})_{\scriptscriptstyle W} (\br,\bp)          \q , \\
     \tau_{q,\mu\nu}^{(s)} (\br) 
		              & = & \, \frac{1}{4} \, \nabla_\mu \nabla_\nu \, \int_{\mathbb{R}^3} \frac{d^3 \vec{p}}{(2\pi\hbar)^3} 
									      \, (\hat{\rho}_{q,\br})_{\scriptscriptstyle W} (\br,\bp) 
		               \, + \, \int_{\mathbb{R}^3} \frac{d^3 \vec{p}}{(2\pi\hbar)^3} \, \frac{p_\mu p_\nu}{\hbar^2} 
											  \, (\hat{\rho}_{q,\br})_{\scriptscriptstyle W} (\br,\bp)          \q , \\
        Q_q (\br) & = & \, \frac{1}{16} \, \Delta \Delta \, \int_{\mathbb{R}^3} \frac{d^3 \vec{p}}{(2\pi\hbar)^3} 
				                \, (\hat{\rho}_{q,\br})_{\scriptscriptstyle W} (\br,\bp) 
		               \, + \, \int_{\mathbb{R}^3} \frac{d^3 \vec{p}}{(2\pi\hbar)^3} \, \frac{p^4}{\hbar^4} 
									      \, (\hat{\rho}_{q,\br})_{\scriptscriptstyle W} (\br,\bp)
                   \, - \, \frac{1}{2} \, \Delta \, \int_{\mathbb{R}^3} \frac{d^3 \vec{p}}{(2\pi\hbar)^3} \, \frac{p^2}{\hbar^2} 
									      \, (\hat{\rho}_{q,\br})_{\scriptscriptstyle W} (\br,\bp)           \nn \\ 
									& \phantom{=} & \hskip 8 cm + \, \int_{\mathbb{R}^3} \frac{d^3 \vec{p}}{(2\pi\hbar)^3} 
									      \, \frac{(\bp \cdot \vnabla)^2}{\hbar^2} 
							          \, (\hat{\rho}_{q,\br})_{\scriptscriptstyle W} (\br,\bp)              \q , \\
								    &   &                                                                  \nn \\					
	J_{q,\mu\nu}(\br) & = & \, \int_{\mathbb{R}^3} \frac{d^3 \vec{p}}{(2\pi\hbar)^3} 
	                        \, \frac{p_\mu}{\hbar} \, (\hat{s}_{q,\br,\nu})_{W} (\br,\bp)   \q , \\ 
 C_{q,\mu\nu} (\br) & = & \, \frac{1}{4} \, \Delta \int_{\mathbb{R}^3} \frac{d^3 \vec{p}}{(2\pi\hbar)^3} 
                          \, \frac{p_\mu}{\hbar} \, (\hat{s}_{q,\br,\nu})_{W} (\br,\bp) 
									   \, - \, \int_{\mathbb{R}^3} \frac{d^3 \vec{p}}{(2\pi\hbar)^3} \, \frac{\bp^2}{\hbar^2} 
									        \, \frac{p_\mu}{\hbar} \, (\hat{s}_{q,\br,\nu})_{W} (\br,\bp)    \nn \\ 
										& \phantom{=} & \hskip 8 cm - \, \frac{1}{2} \, \nabla_\mu \int_{\mathbb{R}^3} \frac{d^3 \vec{p}}{(2\pi\hbar)^3} 
									        \, \frac{(\bp \cdot \vnabla)}{\hbar} 
				   						    \, (\hat{s}_{q,\br,\nu})_{W} (\br,\bp)                          \q , 
\eqr
with 
\begin{align}
             (\hat{\rho}_{q,\br})_{\scriptscriptstyle W} (\br,\bp) 
						          & = \tr_{\sigma} [(\hat{\rho}_q)_{\scriptscriptstyle W} (\br,\bp)]                     \q , \nn \\
             (\hat{\bs}_{q,\br})_{\scriptscriptstyle W} (\br,\bp) 
						          & = \tr_{\sigma} [(\hat{\rho}_q)_{\scriptscriptstyle W} (\br,\bp) \  \hat{\vsigma} ]   \q . \nn
\end{align}
If $H_{q,cl}$ is invariant under rotations in momentum space, the above expressions can be simplified since 
the angular average can be done explicitly. 
By instance, we define, following~\citeRefdot{Gra78}
\begin{equation}
\hat{\alpha}_{q,k} (\br) = \int_0^\infty dp \ p^{k-1-n} 
                      \bigg\langle (\hat{\rho}_{q,\br})_{\scriptscriptstyle W} (\br, \bp) \bigg\rangle  \q ,
\end{equation}
where $\langle f (\vec{r},\vec{p}) \rangle = \frac{1}{4\pi} \int_{S^2} d^2(\frac{\vec{p}}{p}) \  f(\vec{r},\vec{p})$ is the momentum angular average over the unit 2-sphere $S^2$. 
Taking $H_{q,cl}$ as the new integration variable~\footnote{We have to be careful at N2LO with this change of variable, since, for negative values of the $p$ highest order term, the energy density starts to quickly decrease to $-\infty$ for large values of $p$. This generates a high momentum unphysical set for which the associated energies are below the Fermi level. One may need to introduce a momentum cutoff at the N2LO level to avoid this unphysical region.}, we can then write	
\begin{eqnarray}
\hat{\alpha}_{q,k} (\br) =  \frac{p_F^k}{k}\mathbb{1}_\sigma - \sum_{r=1}^\infty \frac{1}{r!}\left[ \left(-\frac{1}{\frac{\partial H_{q,cl}}{\partial p}}\frac{\partial}{\partial p}\right)^{r-1}\left(\frac{ p^{k-1} }{{\frac{\partial H_{q,cl}}{\partial p}}} \langle{\cal \hat{G}}_{q,r}\rangle\right)\right] \Bigg|_{p=p_F(\br)}
\end{eqnarray}
Because of the spin-orbit and N2LO terms, we also need
\begin{eqnarray}
\hat{\alpha}_{q,k,\mu} (\br) = - \sum_{r=1}^\infty \frac{1}{r!}\left[ \left(-\frac{1}
{\frac{\partial H_{q,cl}}{\partial p}}\frac{\partial}{\partial p}\right)^{r-1}\left(\frac{ p^{k-2} }{\frac{\partial H_{q,cl}}{\partial p}} \langle{p_\mu \; \cal \hat{G}}_{q,r}\rangle\right)\right]  \Bigg|_{p=p_F(\br)}
\end{eqnarray}
\begin{eqnarray}
\hat{\alpha}_{q,k,\mu\nu} (\br) = \frac{p_F^k}{3 k}\mathbb{1}_\sigma \delta_{\mu\nu} - \sum_{r=1}^\infty \frac{1}{r!}\left[ \left(-\frac{1}{\frac{\partial H_{q,cl}}{\partial p}}\frac{\partial}{\partial p}\right)^{r-1}\left(\frac{ p^{k-3} }{\frac{\partial H_{q,cl}}{\partial p}}
\langle{p_\mu p_\nu \; \cal \hat{G}}_{q,r}\rangle \right)\right]  \Bigg|_{p=p_F(\br)}
\end{eqnarray}
We can thus write
\bsub
\bqr
     \rho_q (\br) & = & \, \frac{\tr_\sigma (\hat{\alpha}_{q,3})}{2\pi^2 \hbar^3}               \q , \\ 
     \tau_q (\br) & = & \, \frac{\Delta \tr_\sigma (\hat{\alpha}_{q,3})}{8\pi^2 \hbar^3}
		             \, + \, \frac{\tr_\sigma (\hat{\alpha}_{q,5})}{2\pi^2 \hbar^5}                 \q , \\
		 \tau_{q,\mu\nu}^{(s)} (\br) & = & \, \frac{\nabla_\mu \nabla_\nu \tr_\sigma (\hat{\alpha}_{q,3})}{8\pi^2 \hbar^3}
		             \, + \, \frac{\tr_\sigma (\hat{\alpha}_{q,5,\mu \nu})}{2\pi^2 \hbar^5}           \q , \\
        Q_q (\br) & = & \, \frac{\Delta^2 \tr_\sigma (\hat{\alpha}_{q,3})}{32\pi^2 \hbar^3} 
		          \, - \, \frac{\Delta \tr_\sigma (\hat{\alpha}_{q,5})}{4\pi^2 \hbar^5}
              \, + \, \frac{\tr_\sigma (\hat{\alpha}_{q,7})}{2\pi^2 \hbar^7}
						  \, + \, \frac{\nabla_\mu \nabla_\nu \tr_\sigma (\hat{\alpha}_{q,5,\mu \nu})}{2\pi^2 \hbar^5} \q , \\
								  &   &                                                                                     \nn \\					
	J_{q,\mu\nu}(\br) & = & \, \frac{\tr_\sigma (\hat{\alpha}_{q,4,\mu} \hat{\sigma}_\nu)}{2\pi^2 \hbar^4}   \q , \\ 
 C_{q,\mu\nu} (\br) & = & \, \frac{\Delta \tr_\sigma (\hat{\alpha}_{q,4,\mu} \hat{\sigma}_\nu)}{8\pi^2 \hbar^4} 
									 \, - \, \frac{\nabla_\mu \nabla_\kappa \tr_\sigma (\hat{\alpha}_{q,4,\kappa} \hat{\sigma}_\nu)}{4\pi^2 \hbar^4} 
								   \, - \, \frac{\tr_\sigma (\hat{\alpha}_{q,6,\mu} \hat{\sigma}_\nu)}{2\pi^2 \hbar^6}     \q .
\eqr
\esub
Concerning the NLO densities, our results agree with~\citeRefdot{Gra79}. 
The only differences lie in the new densities and currents specific to the N2LO order.
%
%
\subsection{The one-body N2LO hamiltonian}
\label{sec:one_body_mean_field}
%
%

The form of the hamiltonian we derived earlier is not very well suited for the Wigner-Kirkwood transform.
We thus consider the following form
\begin{eqnarray}
\langle \br | \hat{\tilde{h}}_{q} | \br' \rangle 
             & = & \, \tilde U_q \left(\frac{\br+\brp}{2}\right) 
						       \, \delta(\br-\brp) \hat{\mathbb{1}}_\sigma
                 \, - \, \tilde{B}_{q,\mu\nu} \left(\frac{\br+\brp}{2}\right) 
							     \, \nabla_{\mu} \nabla_\nu \delta(\br-\brp) \hat{\mathbb{1}}_\sigma
 			     \, + \, \tilde D_q \left (\frac{\br+\brp}{2}\right) 
							     \, \Delta \Delta \delta(\vec r- \vec r^\prime) \hat{\mathbb{1}}_\sigma \nn \\
             &   &  
						  \, - \, i \, \hat{\vsigma} \cdot \left[ 
							     \, \tilde{\vec{V}}_{q} \left(\frac{\br+\brp}{2}\right) \times \vnabla \Delta \delta (\br-\brp) 
												                    \, \right] 
							\, - \, i \, \hat{\vsigma} \cdot \left[ 
							    \, \tilde{\vec{W}}_{q}\left(\frac{\br+\brp}{2}\right) \times \vnabla \delta (\br-\brp) 
													                  \, \right] 
\end{eqnarray}  
Both are of course equivalent if their fields verify the relations:
\begin{eqnarray}
 \tilde D_q & = & D_q                                                 \q , \nn \\
\tilde {B}_{q,\mu\nu} & = & B_{q,\mu\nu} + \nabla_\mu \nabla_\nu D_q - \frac{1}{2} \Delta D_q \delta_{\mu\nu}  \q , \nn \\
 \tilde U_q & = & U_q + \frac{1}{4} \nabla_\nu \nabla_{\mu} B_{q,\mu\nu} 
											+ \frac{1}{16} \Delta \Delta D_q                \q , \nn \\
\vec{\tilde{V}_q} & = &\vec{V_q}                                      \q , \nn \\
\vec{\tilde{W}_q} & = &\vec{W_q} - \frac{1}{4} \, \Delta \vec{V_q}    \q ,
\end{eqnarray}
The Wigner transform of $\langle \br | \hat{\tilde{h}}_{q} | \br' \rangle$ is then straightforward : 
\begin{align}
(\hat{h}_q)_{\scriptscriptstyle W} (\vec{R},\bp) & = \tilde{U}_q ( \vec{R} )  \hat{\mathbb{1}}_\sigma
			         \, + \, \tilde{B}_{q,\mu\nu} (\vec{R}) \frac{p_\mu p_\nu}{\hbar^2}  \hat{\mathbb{1}}_\sigma
					 \, + \, \tilde{D}_q(\vec{R}) \frac{\bp^4}{\hbar^4}  \hat{\mathbb{1}}_\sigma                         
                     \, + \, \hat{\vsigma} \cdot \left( \tilde{\bW}_q (\vec{R}) \times \frac{\bp}{\hbar} \right) 
					                    \, - \, \hat{\vsigma} \cdot \left( \tilde{\bV}_q (\vec{R}) \times \frac{\bp}{\hbar} \right) \, 
															          \frac{\bp^2}{\hbar^2} 
\end{align}
As we can see, this hamiltonian is far much more complicated than the ones considered in~\citeRefsdot{Gra78,Gra79}. 
Because of this complexity the calculation is no longer possible formally, which is the main advantage we are looking for. 
In particular, the $\bp^4$ and $p_\mu p_\nu$ terms in the hamiltonian make a very tedious calculation of the $\hat{\alpha}$ 
since their denominator would contain polynomials of $p$ and trigonometric functions so that only a numerical angular 
average becomes feasible. 
Moreover, the N2LO fields in the single particle hamiltonian would lead to a equation over $\tau_{\mu \nu}$ 
highly non linear which is not easily solvable by hand. 
Obviously, approximations are mandatory and we will proceed perturbatively: 
we will keep only NLO potentials in the Wigner transform of the one-body hamiltonian but we will consider the N2LO terms 
in the functional. 
As already described in~\citeRefdot{Becker17}, the 4-gradient terms give small numerical contributions to the densities 
so that the approximation made here seems totally justified. 
In that case, the total one-body hamiltonian simply reduces to:
\begin{equation}
(\hat{h}_q)_{\scriptscriptstyle W} (\bR,\bp) = U_{q} (\bR) \hat{\mathbb{1}}_\sigma
			     \, + \, f_{q} (\bR) \frac{\bp^2}{2m}  \hat{\mathbb{1}}_\sigma
           + \hbar \, \hat{\vsigma} \cdot \left( \bg_{q} (\bR) \times \bp \right)
\end{equation}
with a direct correspondence with~\citeRefdot{Gra79}:
\begin{eqnarray}
                 U_{q} & = & \, \tilde U_q \, = \, U_q \, + \, \dfrac{1}{4}\Delta B_q               \q , \\
                 f_{q} & = & \, \frac{2m}{\hbar^2} \, \tilde B_q \, = \, \frac{2m}{\hbar^2} \, B_q   \q , \\
               \bg_{q} & = & \, \frac{1}{\hbar^2} \vec{\tilde{W}}_q \, = \, \frac{1}{\hbar^2} \bW_q \q .
\label{eq:gvsW}
\end{eqnarray}
Explicit expressions for $\hat{\cal G}_i$ and $\hat{\alpha}_i$ functions are given in Appendix \ref{app:Gi}. 
They contain several contributions (Thomas-Fermi (TF), Extended Thomas-Fermi at order $\hbar^2$ (ETF2) 
and spin-orbit (ETF2so)) which are easily identified. 
We are thus now in position to determine the different densities and currents within the approximation described above. 
We obtain (the $\br$ dependencies of all the densities are omitted for sake of simplicity)
\bqr
\label{eq:tau_init}
     \tau_q & = & \, \tau_q^{\rm TF} \, + \, \tau_q^{\rm ETF2,c} \, + \, \tau_q^{\rm ETF2,so}
\eqr
with
\bsub
\bqr
       \tau_q^{\text{TF}} & = & \, \frac{3}{5} \, \left( 3 \pi^2 \right)^{2/3} \, \rho_q^{5/3}         \q , \\
\label{eq:tau_rho}
   		  \tau_q^{\rm ETF2,c} & = & \, \frac{1}{3} \, \Delta \rho_q 
						               \, + \, \frac{1}{36} \, \frac{(\nabla \rho_q)^2}{\rho_q} 
						               \, + \, \frac{1}{6} \, \frac{\nabla f_q \cdot \nabla \rho_q}{f_q} 
						               \, - \, \frac{1}{12} \, \rho_q \, \left( \frac{\nabla f_q}{f_q} \right)^2
						               \, + \, \frac{1}{6} \, \rho_q \, \frac{\Delta f_q}{f_q}                     \q , \\  
	     \tau_q^{\rm ETF2,so} & = & \, \frac{1}{2} \, \left( \frac{2m}{\hbar^2} \right)^2
			                                         \, \rho_q \, \left( \frac{\bW_q}{f_q} \right)^2         \q , 
\eqr
\esub
and similarly for the other densities or currents
\bsub
\label{eq:tau_func}
\bqr
       \tau_{q, \mu \nu}^{(s),\text{TF}}
		                     & = & \, \frac{1}{5} \, \left( 3 \pi^2 \right)^{2/3} 
												                      \, \rho_q^{5/3} \, \delta_{\mu\nu}      \q , \\    												
	 \tau_{q, \mu\nu}^{(s),ETF2,c} & = & \, \frac{1}{3} \, \delta_{\mu\nu} 
											                   \, \left\{ \frac{1}{6} \Delta \rho_q 
													                        + \frac{1}{6} \frac{\nabla f_q \cdot \nabla \rho_q}{f_q} 
													               \, \right\}                                   \nn \\ 
																	&   &																											
		                  +  \frac{1}{6} \, \nabla_\mu \nabla_\nu \rho_q 
											+  \frac{1}{36} \, \frac{(\nabla_\mu \rho_q) (\nabla_\nu \rho_q)}{\rho_q} 
											+  \rho_q \, \left\{ \frac{1}{6} \, \frac{\nabla_\mu \nabla_\nu f_q}{f_q} 
											                   - \frac{1}{12} \, \frac{(\nabla_\mu f_q) (\nabla_\nu f_q)}{f_q^2}
															 \right\}                                               \q , \\
 \tau_{q, \mu\nu}^{(s),ETF2,so} & = & \, \frac{1}{4} \, \left( \frac{2m}{\hbar^2} \right)^2\
	                                      \, \rho_q \, \left\{ \, \left( \frac{\bW_q}{f_q} \right)^2
	                                                       \, \delta_{\mu\nu}  
	                                                  \, - \, \frac{W_{q,\mu} W_{q,\nu}}{f_q^2}  
																							           \, \right\}                  \q , \\
											   &   &                                                         \nn \\						
         Q_q^{\text{TF}} & = & \, \frac{3}{7} \, \left( 3 \pi^2 \right)^{4/3} 
				                       \, \rho_q^{7/3}                                        \q , \\
        Q_{q}^{\rm ETF2,c} & = & \, \frac{5}{3} \, \tau_q^{\rm TF} 
				                       \, \left\{ \, - \, \frac{1}{6} \, \left( \frac{\nabla f_q}{f_q} \right)^2
											                    \, + \, \frac{1}{3} \, \frac{\Delta f_q}{f_q} 
																			    \, + \, \frac{5}{54} \, \left( \frac{\nabla \rho_q}{\rho_q} \right)^2
																			    \, + \, \frac{1}{3} \, \frac{\nabla f_q \cdot \nabla \rho_q}{f_q \rho_q} 
																			    \, - \, \frac{1}{9} \, \frac{\Delta \rho_q}{\rho_q} 
														   \, \right\}                                             \q , \\
			 Q_{q}^{\rm ETF2,so} & = & \, \frac{5}{3} \, \left( \frac{2m}{\hbar^2} \right)^2
			                                        \, \tau_q^{\rm TF}           
																							\, \left( \frac{\bW_q}{f_q} \right)^2    \q ,
\eqr
\esub
Concerning the currents, there is no TF contribution and we have
\bsub
\bqr
\label{eq:JC_vec}
 		  \bJ_{q}^{\rm ETF2} & = & \, - \, \left( \frac{2m}{\hbar^2} \right) 
			                              \, \rho_q \, \frac{\bW_q}{f_q}           \q , \\
		  \bC_{q}^{\rm ETF2} & = & \, \frac{5}{3} 
			                         \, \left( \frac{2m}{\hbar^2} \right)
			                         \, \tau_q^{\rm TF} \, \frac{\bW_q}{f_q}        \q .
\eqr
\esub
One can easily check the agreement of our results with~\citeRefdot{Bra85} and~\citeRefsdot{Gra78,Gra79} 
for all equations concerning NLO currents and densities. 	
Coming back to the full N2LO one-body hamiltonian one can justify the "perturbative treatment" 
of the N2LO component by inspecting by instance~\citeqdot{eq:tau_rho}:
at N2LO order, the $f$ field depends on $\tau$ itself, so that
the extraction of $\tau$ from the highly non linear~\citeqdot{eq:tau_init} would be impossible.
This remark holds also for the other densities.
											
%
%
\section{The surface energy in 1D semi-infinite nuclear matter}
\label{sec:E_S}
%
%
We use here the definition of the surface energy extracted from the simple one-dimensional semi-infinite nuclear matter model which was originally developed by Swiatecki~\cite{Swiatecki51} 
and by Myers and Swiatecki~\cite{MS69,MS74}. 
One considers a medium where the density is constant along the two infinite directions $x$ and $y$ 
and a plane surface perpendicular to the $z-$direction with neutron and proton density profiles denoted 
as $\rho_q(z)$ ($q=n,p$).
Since we are only interested by the surface energy of a symmetric system we consider here only the isoscalar
density $\rho_0(z) = \rho_n(z) + \rho_p(z)$.
Inside the matter, i.e. for $z \rightarrow - \infty$, $\rho_0(z) \rightarrow \rho_{\rm sat}$ where 
$\rho_{\rm sat}$ is the equilibrium density in infinite nuclear matter and ${\cal E}(z)/\rho_0 \rightarrow a_v$ 
the energy per particle in symmetric infinite nuclear matter. 
Outside the matter, i.e. for $z \rightarrow + \infty$, $\rho_0(z) \rightarrow 0$ and ${\cal E}(z) \rightarrow 0$. 

From now on, the surface energy can be simply written as
\bqr
\label{eq:esurf}
                 E_S & = & \, S \, \int_{-\infty}^{+\infty} \, dz
								                \, \left\{ \, {\cal {E}} \left[ \rho_0 (z) \right] 
																      \, - \, \frac{{\cal {E}} \left[ \rho_{\rm sat} \right]}{\rho_{\rm sat}} \, \rho_{0} (z)
																\, \right\}                                                                \q ,
\eqr

Densities and currents have to respect the same symmetries as our system which are:
\begin{itemize}
\item translation invariance with respect to any axis which is orthogonal to $z$ axis 
\item time translation invariance
\item rotational invariance with respect to $z$ axis
\item reflection invariance with respect to any axis which is orthogonal to $z$ axis 
\item time reversal invariance
\end{itemize}

Space and time translation invariances imply that the densities and currents will not depend on $x,y$ cartesian coordinates and time. 
Time reversal invariance associated with time translation invariance imply a vanishing 
time odd part of the functional as already said. 
Rotational and reflection invariances coupled with translation invariance induce the cancellation  
of every remaining densities and currents components except
$\rho_q$, $\tau^{(s)}_{q,xx}$, $\tau^{(s)}_{q,yy}$ ,$\tau^{(s)}_{q,zz}$, $J_{q,xy}$, $J_{q,yx}$, $Q_q$, $C_{q,xy}$ and $C_{q,yx}$. 
Furthermore, because of rotational invariance again, 
we also have $\tau^{(s)}_{q,xx}= \tau^{(s)}_{q,yy}$, $J_{q,xy}=-J_{q,yx}$ and $C_{q,xy}=-C_{q,yx}$. 
Thus, only the $z$ component of the vector parts of $J_{q,\mu \nu}$ and $C_{q,\mu \nu}$ tensors remains. 
Moreover, since we are here interested in symmetric nuclear matter, we have (omitting the $z$ dependence) :
\bsub
\bqr
      U_n & = & \, U_p \, = \, U                                                                          \q , \\
      f_n & = & \, f_p \, = \, f \, = \, 1 \, + \, \frac{2m}{\hbar^2} \, C_0^\tau \, \rho_0               \q , \\
		\bW_n & = & \, \bW_p \, = \, \bW \, = \, - \, C_0^T \, \bJ_0 \, - \, C_0^{\nabla J} \, \vnabla \rho_0 \q . 
\eqr
\esub
Therefore, the energy density finally takes the following form
\bqr
\label{eq:full_EDF_sym}
\CE  [\rho_{0}] & = & \, \frac{\hbar^2}{2m} \, \tau_{0}
      \, + \, C_0^\rho [\rho_{0}]  \, \rho_{0}^2 
      \, + \, C_0^{\Delta \rho} \, \rho_{0} \, \partial_z^2 \rho_{0}                         
	    \, + \, C_0^\tau \, \rho_{0} \, \tau_{0} 
			\, - \, \frac{1}{2} \, C_0^T \, J_{0,z}^2                                     
	    \, + \, C_0^{\nabla J} \, \rho_{0}  \, \partial_z J_{0,z}        \nn \\ 
		            &   &									
		  \, + \, C_0^{(\Delta \rho)^2} \, \left( \partial_z^2 \rho_{0} \right)^2    
      \, + \, C_0^{M \rho} \, \Big[ \, \rho_{0} \, Q_{0}  
											         \, + \, \tau_{0}^2                     
                               \, + \, 2 \, \left( 2 (\tau^{(s)}_{{0,}xx})^2 + (\tau^{(s)}_{{0,}zz})^2  
                                              \, - \, \tau^{(s)}_{{0,}zz} \, \partial_z^2 \rho_{0}    
																				 \, \right) \, \Big]                             \nn \\
								&   &																				
      \, + \, C_0^{MT} \, \Big[ \, 2 \, J_{0,z} C_{0,z}              
                           \, - \, \frac{1}{2} \, J_{0,z} \partial_z^2 J_{0,z} \, \Big]  \q .                     
\eqr
The variational principle can then be applied with a simple Fermi-like density profile
\be
    \rho_{0} (z) \, = \, \frac{\rho_{\rm sat}}{1 + e^{\alpha z}}     \q ,
\ee
where the $\alpha$ parameter is determined in order to minimize the surface energy and $\rho_{\rm sat}$ is the equilibrium density of symmetric nuclear matter.
The result is
\begin{eqnarray}
                 E_S & = & S \, \left[ \, A \, \alpha^3 \, + \, B \, \alpha \, + \, \frac{C}{\alpha} \, \right] \q ,
\end{eqnarray}
where the $A, B, C$ coefficients can be written in such a way that one can easily identify the different contributions of the
Skyrme EDF and the different levels of approximation (TF, ETF2c, ETF2so). The results are presented below, order by order.
%
%
%
\\

For the sake of clarity, we introduced 
$b=\dfrac{2m}{\hbar^2} C_0^\tau \rho_{\rm sat}$, $c= \dfrac{2m}{\hbar^2} \rho_{\rm sat} C_0^T$ and $d=b-c$ in the following sections.

%
%
\subsection{The TF order}
\label{sec:ABC_TF}
%
%
At the TF order the densities occurring in~\citeqdot{eq:full_EDF_sym} are replaced by their TF values,
and the result is
\bsub
\label{eq:ABC_TF}
\bqr			
	    A^{\text{NLO, TF}} & = & \, 0                                                                         \q , \\
      B^{\text{NLO, TF}} & = & \, - \, \frac{1}{6} \, C_0^{\Delta \rho} \, \rho_{\rm sat}^2                     \q , \\			
      C^{\text{NLO, TF}} & = & \, - \, \frac{3}{5} \left(\frac{3\pi^2}{2}\right)^{2/3} \rho_{\rm sat}^{5/3} 
			                              \, \frac{\hbar^2}{2m} \, \frac{3}{2} \, \left[ 1 + \frac{\pi}{3\sqrt{3}} - \log 3 \right] 
														   \, - \, C_0^\rho \, \rho_{\rm sat}^2 - C_0^{\rho^\gamma} \, \frac{3}{2} \, \rho_{\rm sat}^3 \nn \\
                         &   & 
												       \, + \, C_0^\tau \, \frac{3}{5} \, \left( \frac{3\pi^2}{2} \right)^{2/3} \, \rho_{\rm sat}^{8/3} 
														        \, \left( - \frac{21}{10} - \frac{\pi}{2\sqrt{3}} + \frac{3\log 3}{2} \right)  \q , \\
												 &   &                                                                                      \nn \\		
		 A^{\text{N2LO, TF}} & = & \, \frac{1}{30} \, C_0^{(\Delta \rho)^2} \, \rho_{\rm sat}^2                            \q , \\
     B^{\text{N2LO, TF}} & = & \, \frac{3}{44} \, C^{M \rho}_0 \, \left( \frac{3\pi^2}{2} \right)^{2/3} 
		                                                           \, \rho_{\rm sat}^{8/3}                                 \q , \\
     C^{\text{N2LO, TF}} & = & \, C_0^{M \rho} \, \left( \frac{3\pi^2}{2} \right)^{4/3} 
		                                           \, \rho_{\rm sat}^{10/3} \, \frac{3}{5} 
			                         \, \left( - \frac{351}{49} + \frac{2\pi\sqrt{3}}{7} + \frac{18\log 3}{7} \right)    \q ,
\eqr
\esub

The third term of $C^{\text{NLO, TF}}$ coefficient corresponds to the density dependent term.
This one has been chosen as $\rho^\gamma$ with $\gamma = 1$ in order to compare 
our result with \citeRefdot{Gra78} where the autors considered the SIII parametrization. 
However, many other exponents have been widely used in the literature and we give
in~\citeAppendix{app:dd_term}, the expressions for a various panel of values of $\gamma$. 

%
%
\subsection{The Central ETF2 contribution}
\label{sec:ABC_ETF2c}
%
%
Up to $\hbar^2$, the central ETF2 contributions to the $\tau$, $\tau_{\mu\nu}$ and $Q$ densities generate
contributions to the surface energy through the $A, B, C$ coefficients as \footnote{The $B^{\text{NLO, {\rm ETF2,c}}}+B^{\text{NLO, TF}}$ and $C^{\text{NLO, TF}}$ parts corresponds respectively to the B and C coefficient of eq (III.19) of \cite{Gra78}.}
\bsub
\bqr
\label{eq:ABC_ETF2}
       A^{\text{NLO,{\rm ETF2,c }}} & = & \, 0                                                              \q , \\
       B^{\text{NLO, {\rm ETF2,c }}} & = & \, \frac{\hbar^2}{2 m} \, \frac{1}{18} \, \rho_{\rm sat} 
				                      \, + \, C_0^\tau \, \rho^2_{\rm sat} \, \frac{1}{12} 
															     \, \left[ \, - \frac{7}{9} + \frac{1}{b^2} 
																	              - \frac{b+1}{b^3} \, \log(1+b) \, \right]            \q , \\
       C^{\text{NLO, {\rm ETF2,c }}} & = & \, 0                                                               \q , \\
				                    &   &                                                                     \nn \\
      A^{\text{N2LO,{\rm ETF2,c }}} & = & \, C^{M \rho}_0 \, \rho^2_{\rm sat} 
				                          \, \left[ - \frac{7}{1944} - \frac{13}{324 b} - \frac{3301}{7776 b^2} 
																            - \frac{899}{1296 b^3} - \frac{119}{324 b^4}  \right.     \nn \\
														&   &	\left.	  + \left( \frac{17}{81 b^2} + \frac{20}{27 b^3} 
																			 		  + \frac{379}{432 b^4} 
																					  + \frac{119}{324 b^5} \right) \, \log(1+b) \right]       \q , \\
      B^{\text{N2LO, {\rm ETF2,c }}} & = & \, \left( \frac{3\pi^2}{2} \right)^{2/3} \, C^{M \rho}_0 \, \rho^{8/3}_{\rm sat}
			                            \, \left\{ - \frac{7}{44} + \frac{1}{10 b} - \frac{1}{10 b^2} 
																	           + \frac{2}{3 b^3}       \right.                          \nn \\
													  &   &	\left.  	+ \frac{4+b}{3 b^{11/3}} 
																	 \, \left[ \frac{1}{6} \log \left[ \frac{(1+b^{1/3})^2}{1-b^{1/3}+b^{2/3}}\right] 
																							         - \frac{1}{\sqrt{3}} \arctan\frac{2b^{1/3}-1}{\sqrt{3}} 
																											 - \frac{\pi}{6\sqrt{3}}\right] \right\}       \q , \\
      C^{\text{N2LO, {\rm ETF2,c }}} & = & \, 0                                                               \q . 
\eqr
\esub

%

%
%
\subsection{The Spin-Orbit ETF2 contribution}
\label{sec:ABC_ETF2so}
%
%
We now take into account the spin-orbit field in the one-body hamiltonian. 
This constitutes the most elaborate calculation presented in this paper. 
In our 1D semi infinite nuclear matter model, only the $J_z (z)$ component
of the vector part of the current $\bJ (\br)$ survives.
Combining the ETF2 expression for $\bJ$ (see~\citeqdot{eq:JC_vec}
and the expression for $\bW$ (see~\citeqdot{eq:so_fields})
we can write
\bqr
       \bW & = & \, C_0^T \, \left( \frac{2m}{\hbar^2} \right) \, \rho_0 \, \frac{\bW}{f} \, - \, C_0^{\nabla J} \vnabla \rho_0  \q ,
\eqr 
which gives easily the result already obtained 
in~\citeRefsdot{Bartel08,Jod16}:~\footnote{Notice the misprint in Eq.(20) of~\citeRefdot{Jod16} where $\rho_0$ is missing.}
\bqr
       \bW & = & \, \frac{- C_0^{\nabla J} \, \vnabla \rho_0}{1 - \left( \frac{2m}{\hbar^2} \right) C_0^T \frac{\rho_0}{f}} \q ,
\eqr
As it can be easily seen, we also retrieve the case of \citeRefdot{Gra78} by taking $C_0^T = 0$. 
With the above expression, we can get the spin-orbit contribution for each of the densities and currents:
\bsub
\bqr
\tau_{0, \mu\nu}^{(s),{\rm ETF2,so }} & = & 
  \, \frac{1}{4} \, \left( \frac{2m}{\hbar^2} C_0^{\nabla J} \right)^2
	\, \rho_0 \, \frac{(\vnabla \rho_0)^2 \delta_{\mu\nu} - (\nabla_\mu \rho_0) (\nabla_\nu \rho_0)}{\big(1+d\frac{\rho_0}{\rho_{\rm sat}}\big)^2} \\
               Q_{0}^{\rm ETF2,so } & = & 
	\, \frac{5}{3} \, \left( \frac{2m}{\hbar^2} C_0^{\nabla J} \right)^2 \, \tau_0^{\rm TF}           
	\, \frac{(\vnabla \rho_0)^2}{\big(1+d\frac{\rho_0}{\rho_{\rm sat}}\big)^2}					           									\\	 
             \bJ_{0}^{\rm ETF2,so } & = & \, \left( \frac{2m}{\hbar^2} C_0^{\nabla J} \right) 
			                                  \, \rho_0 \, \frac{\vnabla \rho_0}{1+d\frac{\rho_0}{\rho_{\rm sat}}}      \\
		         \bC_{0}^{\rm ETF2,so } & = & \, - \, \frac{5}{3} \, \left( \frac{2m}{\hbar^2} C_0^{\nabla J} \right)
			                                  \, \tau_0^{\rm TF} \, \frac{\vnabla \rho_0}{1+d\frac{\rho_0}{\rho_{\rm sat}}}
\eqr
\esub
As expected there is no SO contribution at TF order so that the new contributions to the surface energy 
arising from the SO term in the one-body hamiltonian read for NLO parts
\bsub
\bqr
\label{eq:ABC_NLO_ETF2so}
        A^{\text{NLO,{\rm ETF2,so }}} & = & \, 0                                   \q , \\
				B^{\text{NLO, {\rm ETF2,so }}} & = & \, \left( C_0^{\nabla J} \right)^2 
				                            \, \rho^3_{\rm sat} \, \frac{m}{\hbar^2}
				                            \, \frac{1}{d} \, \left[ \frac{1}{d^2} + \frac{1}{2d} - \frac{1}{6} 
																		                        - \left( \frac{1}{d^2} + \frac{1}{d^3} \right) \log(1+d) 
																									 \, \right]              \q , \\
        C^{\text{NLO, {\rm ETF2,so }}} & = & \, 0                                   \q , 
\eqr
\esub
where the $J_{\mu \nu}$ density contributes via the $d$ coefficient.
One can again retrieve the results of \citeRefdot{Gra79} by taking $C_0^T = 0$ 
i.e. $c=0$.~\footnote{The reader may note differences between our result and the Eq. (III.26) in~\citeRefdot{Gra79} due 
to some misprints in the power of the coupling constant $C_0^{\nabla J}$ and in the $\log(1+b)$ factor.}
Concerning for the N2LO contributions, we obtain
\bsub
\bqr
\label{eq:ABC_N2LO_ETF2so}
      A^{\text{N2LO, {\rm ETF2,so }}} & = & 
	   \, C_0^{M\rho} \, \frac{m^4}{\hbar^8} \, \left( \frac{4}{3} \, C_0^{\nabla J} \right)^4 \, \rho^6_{\rm sat} 
			              \, \left[- \frac{567}{4d^8} - \frac{6237}{32d^7} - \frac{4239}{64d^6} 
										         - \frac{81}{32d^5} + \frac{81}{640d^4}    \right.                     \nn \\  
												 	   &   & \hskip 4.3 true cm  \left.  
					  								 + \left( \frac{567}{4d^9} + \frac{8505}{32d^8} + \frac{1215}{8d^7} 
																		+ \frac{405}{16d^6} \right) \, \log(1+d) \right]               \nn \\
                             &   & 
\, + \, C_0^{M\rho} \, \frac{m^2}{\hbar^4} \, \left( \frac{4}{3} \, C_0^{\nabla J} \right)^2 \, \rho^4_{\rm sat} 
                    \, \frac{1}{c^3} \, \left[- \frac{3}{8b^4} - \frac{9}{8b^3} - \frac{9}{8b^2} - \frac{3}{8b} 
																		+ c \left( \frac{1}{4b^5} + \frac{15}{16b^4} + \frac{9}{8b^3} 
																		+ \frac{7}{16b^2} \right) \right] \, \log(1+b)                 \nn \\
                             &   &
\, + \, C_0^{M\rho} \, \frac{m^2}{\hbar^4} \, \left( \frac{4}{3} \, C_0^{\nabla J} \right)^2 \, \rho^4_{\rm sat}
                    \, \frac{1}{c^3} \, \left[ \frac{3}{8b^3} + \frac{15}{16b^2} + \frac{11}{16b} 
										                + c \left( -\frac{1}{4b^4} - \frac{13}{16b^3} - \frac{71}{96b^2} 
																		- \frac{1}{8b} \right) \right]                                 \nn \\
                             &   &  
\, + \, C_0^{M\rho} \, \frac{m^2}{\hbar^4} \, \left( \frac{4}{3} \, C_0^{\nabla J} \right)^2 \, \rho^4_{\rm sat}
                    \, \frac{1}{c^3} \, \left[ \frac{3}{8d^4} + \frac{9}{8d^3} + \frac{9}{8d^2} + \frac{3}{8d} 
										                + c \left( -\frac{7}{4d^5} - \frac{69}{16d^4} - \frac{27}{8d^3} 
																		- \frac{13}{16d^2} \right)    \right.                          \nn \\ 
														 &   &  \hskip 3 true cm  \left.
															      + c^2 \left( \frac{5}{d^6} + \frac{21}{2d^5} + \frac{27}{4d^4} 
																		                           + \frac{5}{4d^3} \right)
																		+ c^3 \left( \frac{39}{2d^7} + \frac{335}{8d^6} + \frac{28}{d^5} 
																		+ \frac{45}{8d^4} \right) \right] \, \log(1+d)                 \nn \\
                             &   &  
\, + \, C_0^{M\rho} \, \frac{m^2}{\hbar^4} \, \left( \frac{4}{3} \, C_0^{\nabla J} \right)^2 \, \rho^4_{\rm sat}
                    \, \frac{1}{c^3} \, \left[ - \frac{3}{8d^3} - \frac{15}{16d^2} - \frac{11}{16d} 
										                + c \left( \frac{7}{4d^4} + \frac{55}{16d^3} + \frac{173}{96d^2} 
																		                          + \frac{1}{8d} \right)   \right.     \nn \\
                             &   &  \hskip 3 true cm \left.
															      + c^2 \left( - \frac{5}{d^5} - \frac{8}{d^4} - \frac{19}{6d^3} - \frac{1}{8d^2} \right)
																		+ c^3 \left( - \frac{39}{2d^6} - \frac{257}{8d^5} - \frac{217}{16d^4} 
																		- \frac{17}{24d^3} + \frac{23}{480d^2} \right) \right]         \nn \\
                             &   & 
\, + \, C_0^{Ms} \, \frac{m^2}{\hbar^4} \, \left( \frac{4}{3} \, C_0^{\nabla J} \right)^2 \, \rho^4_{\rm sat} 
                 \, \left[ \frac{18}{d^6} + \frac{99}{4d^5} + \frac{57}{8d^4} + \frac{3}{80d^2} - \left( \frac{18}{d^7} 
								                          + \frac{135}{4d^6} + \frac{18}{d^5} 
																					+ \frac{9}{4d^4} \right) \, \log(1+d) \right]           \q , \\
      B^{\text{N2LO, {\rm ETF2,so }}} & = & \, \left( C_0^{Ms} - C_0^{M\rho} \right) \, \frac{m^2}{\hbar^4} 
			                              \, \left(\frac{3 \pi^2}{2}\right)^{2/3} \, \left( \frac{4}{3} \, C_0^{\nabla J} \right)^2 
			                              \, \rho^{14/3}_{\rm sat} 
																		\, \left\{ - \frac{63}{2d^5} - \frac{243}{20d^4} + \frac{81}{40d^3} 
															                 - \frac{81}{176d^2}  \right.                        \nn \\
														 &   &	\left.		 - \frac{9(14+11d)}{2d^{17/3}} 
																							   \, \left[ \frac{1}{6} \log \left[ \frac{(1+d^{1/3})^2}{1-d^{1/3}+d^{2/3}} \right] 
																							 - \frac{1}{\sqrt{3}} \arctan\frac{2d^{1/3}-1}{\sqrt{3}} - \frac{\pi}{6\sqrt{3}} \right]
																				\, \right\}                                               \q , \\
      C^{\text{N2LO, {\rm ETF2,so }}} & = & \, 0                                                           \q .
\eqr
\esub
%

%
%
\section{Results and discussion}
%
%
From a practical point of view, we first apply the variational principal to $E[\rho_0]/A$ and $E_S$ 
to determine $\rho_{\rm sat}$ and $\alpha$ respectively. 
The surface energy is then calculated for a surface $S$ given by $S = 4 \pi r^2_{\rm sat}$ where $r_{\rm sat}$ 
is the radius of a nucleon at saturation density defined as $\frac{4}{3}\pi r^3_{\rm sat}\rho_{\rm sat} = 1$. 
In other words, we have $S \equiv 4 \pi (3/4\pi \rho_{\rm sat})^{2/3}$. 
The results are presented in Table~\ref{tab:N2LO} for the SN2LO1 parametrization of~\cite{Becker17}. 
As expected, $\rho_{\rm sat}$ is independent of the level of approximation since N2LO terms do not 
contribute to $E[\rho_0]/A$ in infinite nuclear matter (gradients terms). 
On the contrary, $\alpha$ is slightly modified when the $\hbar^2$ second order-expansion is taken into account, 
in agreement with Grammaticos and Voros~\cite{Gra78}. 
This translates into a difference of 1.1 MeV of the surface energy, which is thus the typical order of magnitude  
we can expect when we go from TF approximation to order $\hbar^2$ in the expansion. 
Actually, the main effect comes from the spin-orbit combined N2LO terms in the functional: we obtain a decrease 
of ~3 MeV, which is of course not negligible. 
However, beyond the numbers, what is really important and motivated the present work, is the possible use of 
the explicit expressions of the surface energy written above, directly in the fitting protocol. 
In that case, what is important is the shift between the "exact" value (solution of HF equations) and the 
approximate value coming from the semi-classical expansion presented here: a constant shift would enable us to fine 
tune directly the surface energy during the fitting procedure, disregarding the numerical value itself. 
Since there is only one stable N2LO parametrization in the literature, we decided to compare with results 
coming from series of Skyrme NLO interactions.

From a practical point of view, we first apply the variational principal to $E[\rho_0]/A$ and $E_S$ 
to determine $\rho_{\rm sat}$ and $\alpha$ respectively. 
The surface energy is then calculated for a surface $S$ given by $S = 4 \pi r^2_{\rm sat}$ where $r_{\rm sat}$ 
is the radius of a nucleon at saturation density defined as $\frac{4}{3}\pi r^3_{\rm sat}\rho_{\rm sat} = 1$. 
In other words, we have $S \equiv 4 \pi (3/4\pi \rho_{\rm sat})^{2/3}$. 
In this way we used three series of NLO Skyrme parametrizations:
i) The SLy5sx series~\cite{Jodon16,Ryssens19} which has been constructed to check the simple
MTF (Modified Thomas-Fermi) semi classical approach in a fitting protocol to adjust the surface 
properties of a symmetric semi infinite nuclear matter system.
The MTF expansion was developed originally in \cite{Krivine79} to simulate effective 
mass and $\hbar^4$ effects with only two gradient terms beyond the TF term of the kinetic energy density.
The authors used such an approximation to generate a simple formula for the 
surface energy in semi infinite nuclear matter~\cite{Treiner86}.
This method was recently generalized to Skyrme EDF containing tensor terms as well as 
a full three-body interaction \cite{Jodon16};
ii) The SV series~\cite{Klupfel09} which vary the bulk properties of infinite matter keeping constants
the others;
iii) The SAMi series~\cite{RocaMaza12} which uses a quite similar protocol based on the SAMi parametrization.
The results are depicted on Fig.~\ref{fig:asurf_SLy5s} for the series of SLy5sx parametrizations,
on Fig.~\ref{fig:asurf_Klupfel} for the series of SV and finally on Fig.~\ref{fig:asurf_SAMi} for the
series of SAMi parametrizations. 
On Fig.~\ref{fig:asurf_SLy5s}, is also shown the results obtained for the series: SIII, SkM, SkM*, SLy5 and SLy5*
parametrizations taken here as references.
On each Figure are presented (left panel) the numerical value of the surface energy obtained 
with an HF solver (empty circles), within the ETF2 approximation presented here (empty triangles),
within the ETF4 approximation described in~\citeRefdot{Gra78} (empty squares) which corresponds to an expansion 
up to $\hbar^4$ of the density matrix and within the MTF (Modified Thomas-Fermi) approximation.
To clearly exhibit the difference of each semi classical calculation we present on right panel of each Figure
the difference with the HF result. 
On right panel of each Figure, we clearly see, for ETF4 and MTF approximations, a constant shift 
with respect to the HF results in SLy5s series (respectively +0.5 and -0.5 MeV). 
Moreover, this shift is actually quite independent of the fitting protocol (see SLy5, SLy5*, SkM, SkM*, 
as well as the SLy5sx, SV and SAMi series). And this is an important feature.
This fact was already noticed in~\citeRefdot{Jodon16} and gives a good insight about the existence of a constant shift: 
the surface energy calculated by hand is sufficient to obtain the result of a full numerical solution of the HF equations. 
The SIII parameter set was added for completeness and does not sustain the previous conclusion. 
However, SIII is a particular case since this parametrization has special features like the density-dependent term by instance. 
If we now come to the ETF2 approximation, the shift, instead of being constant, is seen varying in a range [0.7, 1.2] MeV, which is however quite reasonable.
The ETF2 calculation gives then a quite accurate order of magnitude of the surface energy, which is sufficient to be incorporated in a fitting protocol including the surface energy as a primary ingredient. 
Finally, an important feature of our results comes from the comparison to the HF result (see~\citeRefdot{Bender22}) for the SN2LO1 parametrization 
which exhibits quite the same shift (relative sign and amplitude). 
Even if the statistics if of course not significant, there is however a valuable hint about the possibility to use safely our formulas 
in the fitting procedure of future N2LO parametrizations.

%
%
\begin{table}[h]
\bc
\begin{tabular}{c|c|c|c}
\hline \hline \noalign{\smallskip}
    order  &  TF  &  TF + ETF2,c  &  ETF2 \\[0.3mm]
\noalign{\smallskip}  \hline \noalign{\smallskip}
             $(\hat{h})_{\scriptscriptstyle W}$ 
	         & $\big( f \frac{\bp^2}{2m} + V \big) \, \hat{\mathbb{1}}_\sigma$ 
					 & $\big( f \frac{\bp^2}{2m} + V \big) \, \hat{\mathbb{1}}_\sigma$  
					 & $\big( f \frac{\bp^2}{2m} + V \big) \, \hat{\mathbb{1}}_\sigma 
			\, + \, \hbar \, (\bg \times \bp) \cdot \hat{\vsigma}$                \\[0.3mm]
\noalign{\smallskip}  \hline \noalign{\smallskip}
$\rho_{\rm sat}$ (fm$^{-3})$  &  0.162 &  0.162 &  0.162                    \\[0.3mm]
$\alpha$ (fm$^{-1})$          &  1.916 &  2.021 &  2.233                    \\[0.3mm]
$E_S$ (MeV)                   & 20.41  & 19.29  & 17.68                     \\[0.3mm]
\noalign{\smallskip} \hline \hline
\end{tabular}
\caption{Numerical results obtained within different approximations with the SN2LO1 
parametrization \cite{Becker17}.
Second column gives the TF results, i.e. $\hbar^0$ order.
Third and fourth columns give respectively the results which include TF + ETF2,c
and TF + full ETF2 (central + spin-orbit contributions, i.e. ETF2c and ETF2so parts.}
\label{tab:N2LO}
\ec
\end{table}
%
%

%
%
%
%

%
\bc
\begin{figure}[htbp]
\vskip -1truecm
\begin{center}
\includegraphics[width=0.8\linewidth,bb=20 260 600 620,clip]{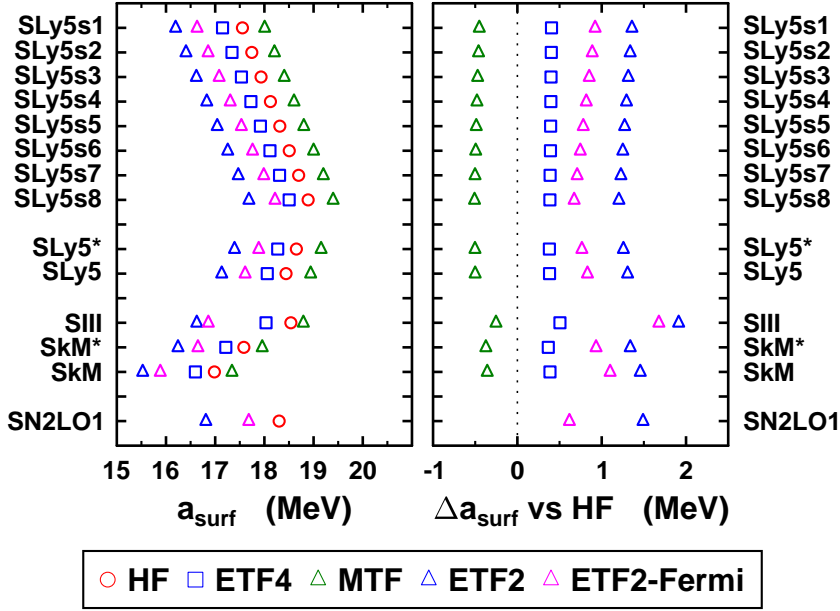} 
\caption{(Color online)
         Surface energy coefficients obtained within HF (empty circles), 
				 MTF (wide triangles) and ETF (empty squares) approaches for the SLy5s 
				 series of Skyrme parametrizations (see text). 
         The Figure has been enriched with the standard SIII, SkM, SkM$^*$,
				 and SLy5 and SLy5$^*$ parametrizations taken as references.
				 The Figure shows also a preliminary HF result for the N2LO
				 SN2LO1 parametrization of~\citeRefdot{Becker17}.}
\label{fig:asurf_SLy5s}
\end{center}
\end{figure}
\ec
%
%
\bc
\begin{figure}[htbp]
\vskip -1truecm
\begin{center}
\includegraphics[width=0.8\linewidth,angle=-90,clip]{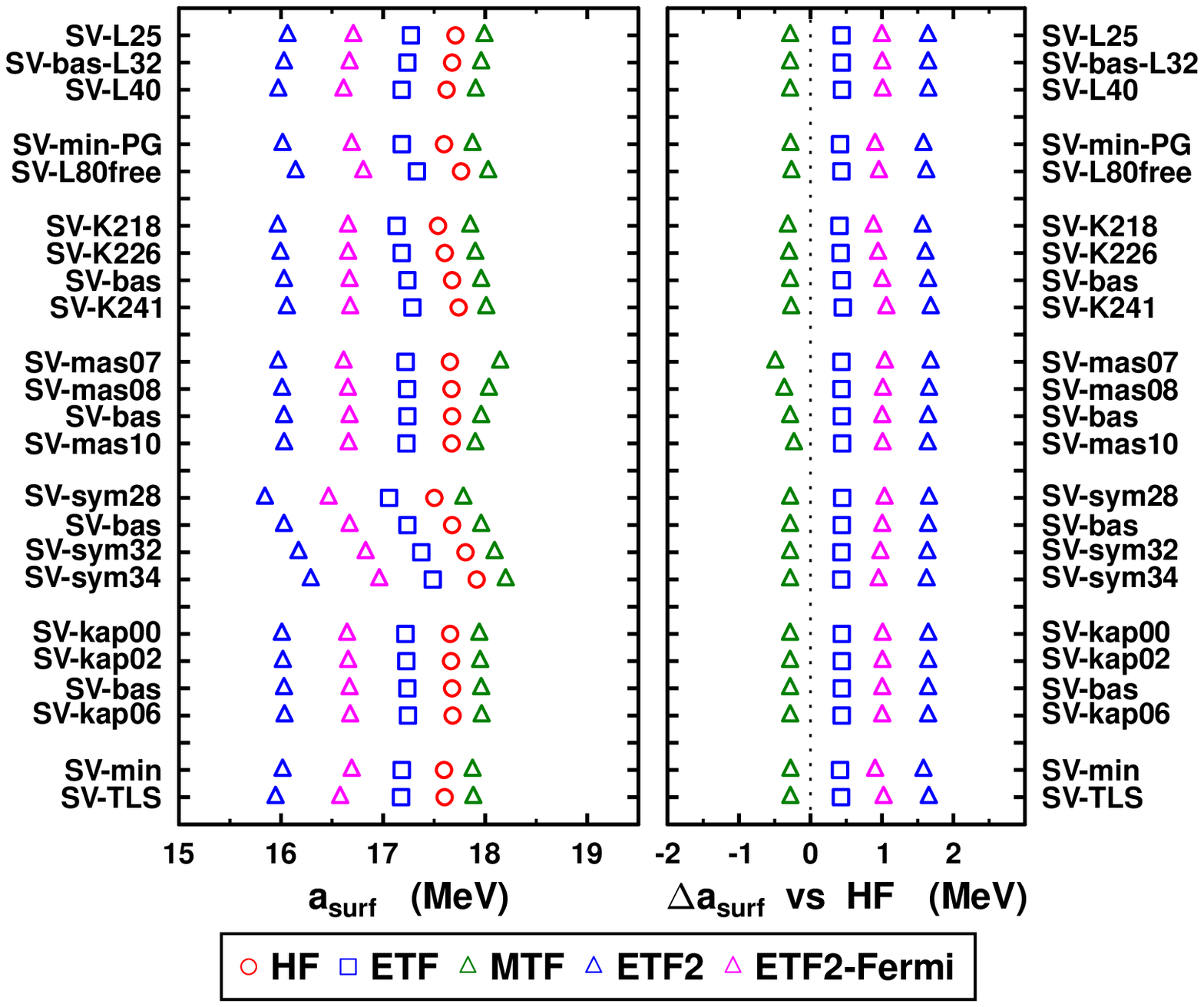} 
\caption{(Color online)
         Same as~\citeFigure{fig:asurf_Klupfel} for the SV series of
				 NLO Skyrme parametrization (see~\citeRefsdot{Klupfel09,PG16}).}
\label{fig:asurf_Klupfel}
\end{center}
\end{figure}
\ec
%

%
\bc
\begin{figure}[htbp]
\vskip -1truecm
\begin{center}
\includegraphics[width=0.8\linewidth,angle=-90,clip]{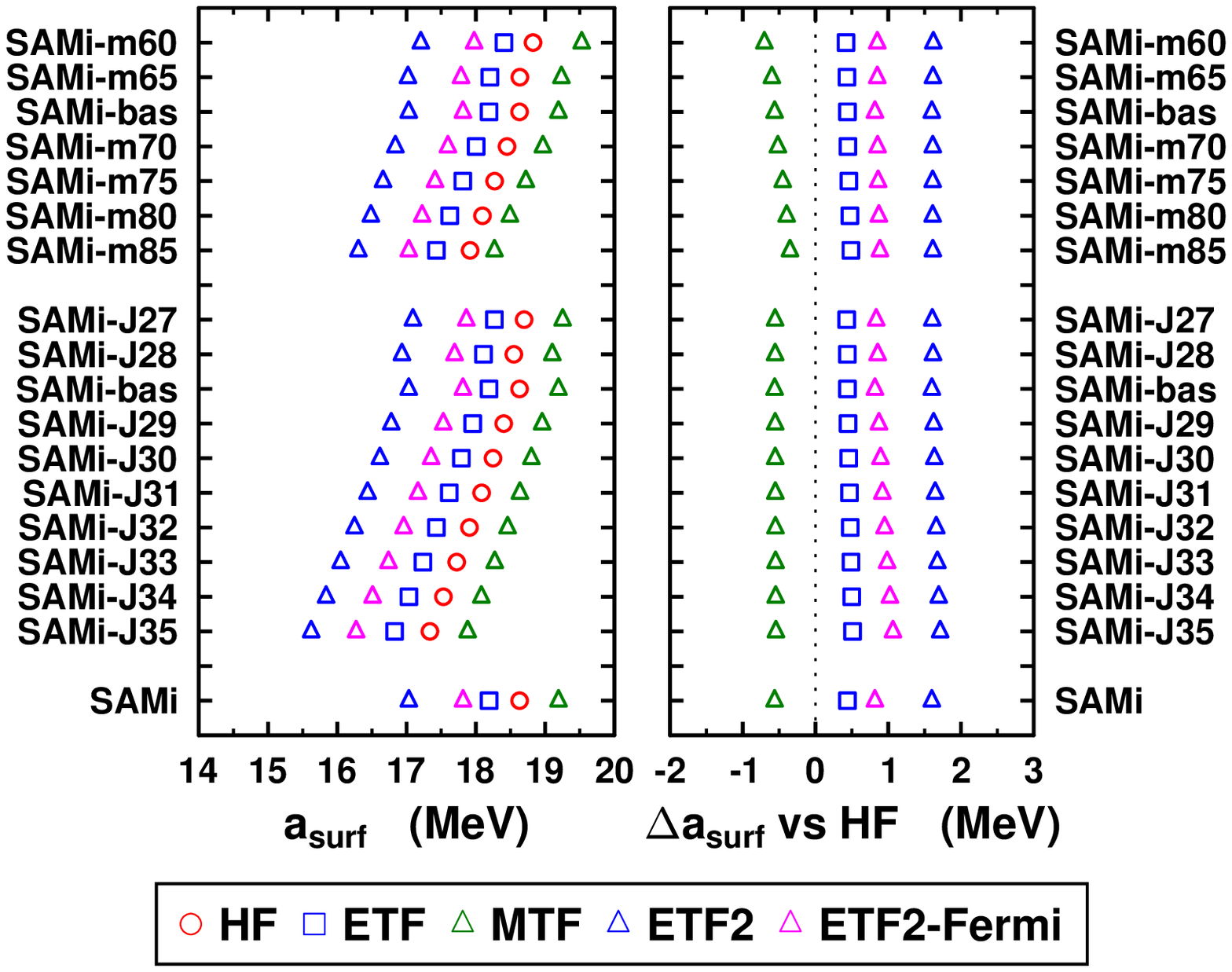} 
\caption{(Color online)
         Same as~\citeFigure{fig:asurf_Klupfel} for the SAMi series of
				 NLO Skyrme parametrization (see~\citeRefsdot{RocaMaza12,Colo16}).}
\label{fig:asurf_SAMi}
\end{center}
\end{figure}
\ec
%
%
%
\section{Conclusion}
%
%

The semiclassical calculations are an alternative to Hartree Fock calculations which allow us to derive observables 
without using wave functions formalism. 
In this paper, we extended the method developed in~\citeRefsdot{Gra78,Gra79} to Skyrme N2LO functionals to get a handly 
expression of the surface energy. 
Our results have been obtained within two approximations. 
The first one is the restriction to the order $\hbar^2$ in the Wigner-Kirkwood expansion of the density matrix. 
This is justified since one of the conclusions of~\citeRefdot{Gra78} is the non necessity to expand the density matrix 
Wigner transform up to $\hbar^4$ order to get a good estimate of the surface energy compared to TF and ETF2 approximations. 
The second approximation lies in neglecting the N2LO terms in the one-body hamiltonian. 
It comes from the impossibility to derive formally the surface energy because of these terms. 
It is however justified by the relative size of N2LO terms which are supposed to be small compared 
to NLO terms (see~\cite{Becker17}).
With this kind of semiclassical approach, there is a loss of around 1 MeV in the precision 
of the surface energy with respect to the Hartree Fock results. 
However, for NLO functionals, the difference between "exact" Hartree Fock and ETF formula is close to constant: 
there is only a constant shift between both results which enable us to incorporate the surface energy directly 
in the fitting procedure.
Since the same behavior appears to be true with the N2LO functional of~\citeRefdot{Becker17}, there is a good hint suggesting the possibility to use the ETF estimations of the surface energy as pseudo-datas for a fitting program of N2LO functionals. 
One can imagine to construct a similar fitting protocol already used in Ref.\cite{Jodon16} to adjust the surface properties
In this regards, the results of this paper open opportunities to fit new N2LO parametrizations which shall 
incorporates new fundamental features by construction.
%

%
%
\section*{Acknowledgments}
%
%

The authors thank M. Bender for very fruitful discussions and for the private communication of the Hartree Fock result for SN2LO1.
The authors thank also P.-G. Reinhard and G. Col\`o for the private
communication of the unpublished parameters of the SV Skyrme 
(SAMi respectively) pseudo-potentials. 
The authors gratefully acknowledge support from the CNRS/IN2P3 Computing 
Center (Lyon-France) for providing computing resources needed for this work.
%

%
%
\begin{appendix}
%
%

%
%
\section{N2LO one body hamiltonian}
\label{app:fields}
%
%
We give here all the expressions of the various parts of the N2LO one body hamiltonian
(see~\citeqdot{eq:hfield}). 
We have successively (omitting the $\br$ dependence of the densities in the right 
hand sides for sake of simplicity)
\bi
    \item central fields
\bsub
\bqr  
      U_q^{\text{NLO}} (\br) & = & \, 2 \, \left[ \, C_0^\rho \, - \, C_1^\rho \, \right] \, \rho_0 
				                      \, + \, 4 \, C_1^\rho \, \rho_q                                 \nn \\
                             &   &
           \, + \, 2 \, \left[ \, C_0^{\rho^{\gamma}} \, - \, C_1^{\rho^{\gamma}} \, \right] 
							       \, \rho_0^{\gamma} \, \rho_0 	   
				   \, + \, 4 \, C_1^{\rho^{\gamma}} \, \rho_0^{\gamma} \, \rho_q   
			     \, +      \, {\gamma} \, \left[ \, C_0^{\rho^{\gamma}} \, + \, C_1^{\rho^{\gamma}} \, \right] 
				                       \, \rho_0^{{\gamma}+1}          
				   \, - \, 4 \, {\gamma} \, C_1^{\rho^{\gamma}} \, \rho_0^{\gamma} \, \rho_q      
				   \, + \, 4 \, {\gamma} \, C_1^{\rho^{\gamma}} \, \rho_0^{{\gamma}-1} \, \rho_q^2    \nn \\		
		                         &   & 
		       \, + \, 2 \, \left[ \, C^{\Delta \rho}_0 \, - \, C^{\Delta \rho}_1 \, \right] \, \Delta \rho_0    
           \, + \, 4 \, C^{\rho}_1 \, \Delta \rho_q                                                                     
		       \, + \, \left[ \, C^{\tau}_0 \, - \, C^{\tau}_1 \, \right] \, \tau_0    
           \, + \, 2 \, C^{\tau}_1 \, \tau_q                                                  \nn \\
		                         &   &			
		       \, + \, \left[ \, C^{\nabla J}_0 \, - \, C^{\nabla J}_1 \, \right] 
		            \, \left( \nabla_\kappa \, J_{\kappa, 0} \right)     
           \, + \, 2 \, C^{\nabla J}_1 
		            \, \left( \nabla_\kappa \, J_{\kappa, q} \right)                             \q , \\
		 U_q^{\text{N2LO}} (\br) & = & 
		            \, 2 \, \left[ \, C_0^{(\Delta \rho)^2} \, - \, C_1^{(\Delta \rho)^2} \, \right] 
		                 \, \Delta \Delta \rho_0
					 \, + \, 2 \, C_1^{(\Delta \rho)^2} \, \Delta \Delta \rho_q                          \nn \\                        
	                           &   &
           \, + \, \left[ \, C_0^{M\rho} \, - \, C_1^{M\rho} \, \right] \, Q_0 
					 \, + \, 2 \, C_1^{M\rho} \, Q_q                                            
           \, - \, 2 \, \left[ \, C_0^{M\rho} \, - \, C_1^{M\rho} \, \right] \, \nabla_\mu \nabla_\nu \tau_{0,\mu\nu}^{(s)} 
					 \, - \, 4 \, C_1^{M\rho} \, \nabla_\mu \nabla_\nu \tau_{q,\mu\nu}^{(s)}           \q , 
\eqr
\esub
where the density dependent coupling constants $C_t^\rho \left[ \rho_0 \right]$ have been expanded as already
noticed in~\citeqdot{eq:full_EDF}. 

    \item effective mass fields 
\bsub
\bqr 
			B_q^{\text{NLO}} (\br) & = & \, \frac{\hbar^2}{2m}                              
			                        \, + \, \left[ \, C^{\tau}_0 \, - \, C^{\tau}_1 \, \right] \, \rho_0              
                              \, + \, 2 \, C^{\tau}_1 \, \rho_q                     \q , \\  
		 B_{q, \mu \nu}^{\text{N2LO}} (\br) & = & 
		                              \, 2 \, \left[ \, C_0^{M \rho} \, - \, C_1^{M \rho} \, \right] \, \tau_0 \delta_{\mu\nu}
		                          \, + \, 4 \, C_1^{M \rho} \, \tau_q \delta_{\mu\nu} 
		                          \, + \, 4 \, \left[ \, C_0^{M \rho} \, - \, C_1^{M \rho} \, \right] 
		                                    \, \tau^{(s)}_{0, \mu \nu}              
																	 \, + \, 8 \, C_1^{M \rho} \, \tau^{(s)}_{q, \mu \nu} 
																	\nn \\ & & \, - \, 2 \, \left[ \, C_0^{M \rho} \, - \, C_1^{M \rho} \right] 
																						 \, \nabla_\mu \nabla_\nu \rho_0 
																	 \, - \, 4 \, C_1^{M \rho} \, \nabla_\mu \nabla_\nu \rho_q   \q , \\
 														  &   &                                                             \nn \\									
      D_q^{\text{N2LO}} (\br) & = & \, \left[ \, C_0^{M \rho} \, - \, C_1^{M \rho} \, \right] \, \rho_0
								               \, + \, 2 \, C_1^{M \rho} \, \rho_q                             \q . 
\eqr
\esub

     \item spin-orbit fields
\bsub
\bqr		
\label{eq:so_fields}
 \bW_q^{\text{NLO}} (\br) & = & \, - \, \left[ \,  C^{\nabla J}_0 \, - \, C^{\nabla J}_1 \, \right] \, \vnabla \rho_0  
			                          \, - \, 2 \, C^{\nabla J}_1 \, \vnabla \rho_q 
                                \, - \, \left[ \, C^T_0 \, - \, C^T_1 \, \right] \, \bJ_0     
		                            \, - \, 2 \, C^T_1 \, \bJ_q                          \q , \\
\bW_q^{\text{N2LO}} (\br) & = & \, 2 \, \left[ \, C_0^{MT} \, - \, C_1^{MT} \, \right] \, \bC_0  
                                \, + \, 4 \, C_1^{MT} \, \bC_q                       \q , \\
	 											  &   &                                                       \nn \\			
\bV_q^{\text{N2LO}} (\br) & = & \, 2 \, \left[ \, C_0^{MT} \, - \, C_1^{MT} \, \right] \, \bJ_0  
                                \, + \, 4 \, C_1^{MT} \, \bJ_q                        \q .
\eqr
\esub
\ei
%

%
%
\section{Explicit expressions of $\hat{\cal G}_i$  and $\hat{\alpha}_i$ functions}
\label{app:Gi}
%

If we use the explicit expressions of $H_{q,cl}$ and $\vec{H_{q,so}}$ with $\vec{x} = (\vec{r},\vec{p})$, we obtain:
\begin{align}
\hat{\cal G}_{q ,0} (\vec{x}, \hbar) & = \hat{\mathbb{1}}_\sigma \nonumber \\
 \hat{\cal G}_{q,1} (\vec{x}, \hbar) & = \hbar \, (\vec{g}_q \times \bp) \cdot \hat{\vsigma}  \nonumber \\
 \hat{\cal G}_{q,2} (\vec{x}, \hbar) & = \frac{\hbar^2}{8m^2} \big[ 8 m^2 \vec{g}_q^2 \bp^2 - 8 m^2 (\vec{g}_q \cdot \bp)^2 + 2 (\bp \cdot \vnabla f_q)^2 - f_q \bp^2 \Delta f_q - 2 m f_q \Delta V_q \big] \hat{\mathbb{1}}_\sigma + {\cal O} (\hbar^3) \nonumber \\
\hat{\cal G}_{q,3} (\vec{x}, \hbar) & = \frac{\hbar^2}{16m^3} \big[ 4 f_q \bp^2 (\bp \cdot \vnabla f_q)^2 + 8 m f_q (\bp \cdot \vnabla f_q) (\bp \cdot \vnabla V_q) - 2 f^2_q \bp^2 (\bp \cdot \vnabla)^2 f_q - 4 m f_q^2 (\bp \cdot \vnabla)^2 V_q - f_q \bp^4 (\vnabla f_q)^2 \nn \\
& \phantom{=} \  - 4 m f_q \bp^2 (\vnabla f_q) \cdot (\vnabla V_q) - 4 m^2 f_q (\vnabla V_q)^2 \big] \hat{\mathbb{1}}_\sigma + {\cal O} (\hbar^3)
\end{align}

We thus get
\begin{small}
\begin{align}
\hat{\alpha}_{q,3} = & \frac{p_F^3}{3} \hat{\mathbb{1}}_\sigma + \hbar^2 \bigg[ p_{F,q} \bigg( \frac{m^2 \vec{g_q}^2}{f_q^2} + \frac{7 (\vnabla f_q)^2}{96 f_q^2} - \frac{\Delta f_q}{12 f_q} \bigg) + \frac{m}{p_{F,q}} \bigg( \frac{(\vnabla f_q) \cdot (\vnabla V_q)}{24 f_q^2} - \frac{\Delta V_q}{12 f_q} \bigg) + \frac{m^2}{p_{F,q}^3} \bigg( - \frac{(\vnabla V_q)^2}{24 f_q^2} \bigg) \bigg] \hat{\mathbb{1}}_{\sigma} \nonumber \\ & + \mathcal{\cal O} (\hbar^3) \\ 
\hat{\alpha}_{q,4,\mu} = & - \hbar \frac{m \ p_{F,q}^3}{3f_q} (\hat \vsigma \times \vec{g_q})_\mu +  \mathcal{\cal O} (\hbar^3) \\ 
\hat{\alpha}_{q,5,\mu \nu} = & \frac{p_{F,q}^5 \delta_{\mu \nu}}{15} \hat{\mathbb{1}}_\sigma + \hbar^2 \bigg[ p_{F,q}^3 \bigg( \frac{2 m^2 \vec{g_q}^2 \delta_{\mu \nu}}{3 f_q^2} - \frac{m^2 g_{q,\mu} g_{q,\nu}}{3 f_q^2} + \frac{19 (\vnabla f_q)^2 \delta_{\mu \nu}}{288 f_q^2} - \frac{(\nabla_\mu f_q) (\nabla_\nu f_q)}{9 f_q^2} - \frac{(\Delta f_q)  \delta_{\mu \nu}}{18 f_q} + \frac{7 (\nabla_\mu \nabla_\nu f_q)}{72 f_q} \bigg) \nonumber \\ & + m \  p_{F,q} \bigg( \frac{\nabla_\mu \nabla_\nu V_q}{12 f_q} - \frac{(\Delta V_q) \delta_{\mu \nu}}{12 f_q} + \frac{(\vnabla f_q) \cdot (\vnabla V_q) \delta_{\mu \nu}}{8 f_q^2} - \frac{(\nabla_\mu f_q)(\nabla_\nu V_q)}{12 f_q^2} - \frac{(\nabla_\nu f_q)(\nabla_\mu V_q)}{12 f_q^2} \bigg) \nonumber \\ & + \frac{m^2}{p_{F,q}} \bigg( \frac{(\vnabla V_q)^2 \delta_{\mu \nu}}{24 f_q^2} \bigg) \bigg] \hat{\mathbb{1}}_\sigma + \mathcal{\cal O} (\hbar^3) \\ 
\hat{\alpha}_{q,5} = & \frac{p_{F,q}^5}{5} \hat{\mathbb{1}}_\sigma + \hbar^2 \bigg[ p_{F,q}^3 \bigg( \frac{5 m^2 \vec{g_q}^2}{3 f_q^2} + \frac{25 (\vnabla f_q)^2}{288 f_q^2} - \frac{5(\Delta f_q)}{72 f_q} \bigg) + m \  p_{F,q} \bigg( \frac{5(\vnabla f_q) \cdot (\vnabla V_q)}{24 f_q^2} - \frac{\Delta V_q}{6 f_q} \bigg) + \frac{m^2}{p_{F,q}} \bigg( \frac{(\vnabla V_q)^2 }{8 f_q^2} \bigg) \bigg] \hat{\mathbb{1}}_\sigma \nonumber \\ & + \mathcal{\cal O} (\hbar^3) \\ 
\hat{\alpha}_{q,6,\mu} = & - \hbar \frac{m \ p_{F,q}^5}{3 f_q} (\hat \vsigma \times \vec{g_q})_\mu + \mathcal{\cal O} (\hbar^3) \\ 
\hat{\alpha}_{q,7} = & \frac{p_{F,q}^7}{7} \hat{\mathbb{1}}_\sigma + \hbar^2 \bigg[ p_{F,q}^5 \bigg( \frac{7 m^2 \vec{g_q}^2}{3 f_q^2} + \frac{7 (\vnabla f_q)^2}{96 f_q^2} \bigg) + m \  p_{F,q}^3 \bigg( \frac{35(\vnabla f_q) \cdot (\vnabla V_q)}{72 f_q^2} - \frac{5(\Delta V_q)}{36 f_q} \bigg) + m^2 p_{F,q} \bigg( \frac{5(\vnabla V_q)^2 }{8 f_q^2} \bigg) \bigg] \hat{\mathbb{1}}_\sigma \nonumber \\ & + \mathcal{\cal O} (\hbar^3) 
\end{align} 
\end{small}
where $p_{F,q}$ is the Fermi momentum. The corresponding Fermi energy is given by $f_q \frac{p_{F,q}^2}{2m} + V_q = E_{F,q}$. Since densities and currents are not space independent, the Fermi momentum and the fields $f_q$ and $V_q$ depend on the space variables unlike $E_{F,q}$. Thus, we get
\begin{align}
m \nabla_\mu V_q & = - \frac{p^2_{F,q} (\nabla_\mu f_q)}{2} - p_{F,q} f_q (\nabla_\mu p_{F,q}) \nn \\
m \nabla_\mu \nabla_\nu V_q & = - \frac{p^2_{F,q} (\nabla_\mu \nabla_\nu f_q)}{2} - p_{F,q} (\nabla_\mu f_q) (\nabla_\nu p_{F,q}) - p_{F,q} (\nabla_\nu f_q) (\nabla_\mu p_{F,q}) \nn \\ & - p_{F,q} f_q (\nabla_\mu \nabla_\nu p_{F,q}) - f_q (\nabla_\mu p_{F,q}) (\nabla_\nu p_{F,q})
\end{align}
so that :
\begin{align}
\hat{\alpha}_{q,3} = & \frac{p_{F,q}^3}{3} \hat{\mathbb{1}}_\sigma + \hbar^2 p_{F,q} \bigg( \frac{m^2 \vec{g_q}^2}{f_q^2} + \frac{(\vnabla f_q)^2}{24 f_q^2} - \frac{\Delta f_q}{24 f_q} + \frac{(\vnabla f_q) \cdot (\vnabla p_{F,q})}{12 f_q p_{F,q}} + \frac{\Delta p_{F,q}}{12 p_{F,q}} + \frac{(\vnabla p_{F,q})^2}{24 p_{F,q}^2} \bigg) \hat{\mathbb{1}}_{\sigma} + \mathcal{\cal O} (\hbar^3) \\ 
\hat{\alpha}_{q,4,\mu} = & - \hbar \frac{m \ p_{F,q}^3}{3f_q} (\hat \vsigma \times \vec{g})_\mu +  \mathcal{\cal O} (\hbar^3) \\ 
\hat{\alpha}_{q,5,\mu \nu} = & \frac{p_{F,q}^5 \delta_{\mu \nu}}{15} \hat{\mathbb{1}}_{\sigma}+ \hbar^2 p_{F,q}^3 \bigg( \frac{2 m^2 \vec{g_q}^2 \delta_{\mu \nu}}{3 f^2_q} - \frac{m^2 g_{q,\mu} g_{q,\nu}}{3 f^2_q} + \frac{(\vnabla f_q)^2 \delta_{\mu \nu}}{72 f^2_q} - \frac{(\nabla_\mu f_q) (\nabla_\nu f_q)}{36 f^2_q} - \frac{(\Delta f_q)  \delta_{\mu \nu}}{72 f_q} + \frac{ \nabla_\mu \nabla_\nu f_q}{18 f_q} \nonumber \\ & - \frac{\nabla_\mu \nabla_\nu p_{F,q}}{12 p_{F,q}} + \frac{(\Delta p_{F,q}) \delta_{\mu \nu}}{12 p_{F,q}} + \frac{(\vnabla f_q) \cdot (\vnabla p_{F,q}) \delta_{\mu \nu}}{12 f_q p_{F,q}} - \frac{(\nabla_\mu p_{F,q})(\nabla_\nu p_{F,q})}{12 p^2_{F,q}} + \frac{(\vnabla p_{F,q})^2 \delta_{\mu \nu}}{8 p_{F,q}^2} \bigg) \hat{\mathbb{1}}_{\sigma} + \mathcal{\cal O} (\hbar^3) \\ 
\hat{\alpha}_{q,5} = & \frac{p^5_{F,q}}{5} \hat{\mathbb{1}}_{\sigma} + \hbar^2 p^3_{F,q} \bigg( \frac{5 m^2 \vec{g_q}^2}{3 f^2_q} + \frac{(\vnabla f_q)^2}{72 f^2_q} + \frac{\Delta f_q}{72 f_q} + \frac{(\vnabla f_q) \cdot (\vnabla p_{F,q})}{4 f_q p_{F,q}} + \frac{\Delta p_{F,q}}{6 p_{F,q}} + \frac{7(\vnabla p_{F,q})^2 }{24 p^2_{F,q}} \bigg) \hat{\mathbb{1}}_{\sigma} + \mathcal{\cal O} (\hbar^3) \\ 
\hat{\alpha}_{q,6,\mu} = & - \hbar \frac{m \ p^5_{F,q}}{3f_q} (\hat \vsigma \times \vec{g_q})_\mu +  \mathcal{\cal O} (\hbar^3) \\ 
\hat{\alpha}_{q,7} = & \frac{p^7_{F,q}}{7} \hat{\mathbb{1}}_{\sigma} + \hbar^2 p^5_{F,q} \bigg( \frac{7 m^2 \vec{g_q}^2}{3 f^2_q} + \frac{5(\Delta f_q)}{72 f_q} - \frac{(\vnabla f_q)^2}{72 f^2_q} + \frac{5(\vnabla f_q) \cdot (\vnabla p_{F,q})}{12 f_q p_{F,q}} + \frac{5(\Delta p_{F,q})}{36 p_{F,q}} + \frac{55(\vnabla p_{F,q})^2 }{72 p^2_{F,q}} \bigg) \hat{\mathbb{1}}_{\sigma} + \mathcal{\cal O} (\hbar^3) 
\end{align}

%
%
\section{Density dependent term in the $C^{\text{NLO, TF}}$ coefficient}
\label{app:dd_term}
%
%
The $C^{\text{NLO, TF}}$ coefficient of the surface energy (see~\citeqdot{eq:ABC_TF}) 
is obviously dependent of the exponent $\gamma$ of the density dependence occurring 
in the Skyrme EDF. 
In the text, i.e. in~\citeqdot{eq:ABC_TF}, we gave the result for $\gamma = 1$.
In the general case, the contribution due to the density-dependent term, namely
\bqr 
 C^{\text{NLO, TF}} & \equiv & \, - \, C_0^{\rho^\gamma} \, \rho_0^3 \, \frac{3}{2}     \q ,
\eqr
has to be replaced by
\bqr 
 C^{\text{NLO, TF}} & \equiv & \, - \, \, C_0^{\rho^\gamma} \, \rho_0^{\gamma+2} \, C_\gamma \q ,
\eqr
where the $C_\gamma$ coefficient can be read in~\citeTable{tab:dd_power} for the density dependencies
commonly used in the literature.		
%
%
\begin{table}[h]
\bc
\begin{tabular}{ccc}
\hline \hline \noalign{\smallskip}
  $\gamma$ && $C_\gamma$                                                   \\[0.3mm]
\noalign{\smallskip}  \hline \noalign{\smallskip}
  $1$      && $ \frac{3}{2}$                                               \\[0.3mm]
  $2/3$    && $ \frac{21}{10} + \frac{\pi}{2\sqrt{3}} - \frac{3\log 3}{2}$ \\[0.3mm]
	$1/2$    && $ \frac{8}{3}   - \log 4$                                    \\[0.3mm] 
	$1/3$    && $ \frac{15}{4}  - \frac{\pi}{2\sqrt{3}} - \frac{3\log 3}{2}$ \\[0.3mm]
	$1/4$    && $ \frac{24}{5}  - \frac{\pi}{2}         - \log 8$            \\[0.3mm]
	$1/5$    && $  \frac{35}{6} - \frac{5\log5}{4} 
	                            - \frac{\sqrt{5}}{4} \log \frac{\sqrt{5}+1}{\sqrt{5}-1} 
	                            - \frac{\pi}{2} \sqrt{1+\frac{2}{\sqrt{5}}}$ \\[0.3mm]
	$1/6$    && $ \frac{48}{7}  - \frac{\pi\sqrt{3}}{2} - \frac{3\log3}{2} 
	                            - \log 4$                                    \\[0.3mm]
\noalign{\smallskip} \hline \hline
\end{tabular}
\caption{Expressions of the $C_\gamma$ coefficient (see text) according to
the value of the power of the density dependent term in the NLO Skyrme EDF.}  
\label{tab:dd_power}
\ec
\end{table}
%
%
%
\end{appendix}
%
%
\bibliography{prc_ETF_N2LO}
%



\end{document}